\documentclass[preprint,aps]{revtex4}
\usepackage{graphicx}
\usepackage{dcolumn}
\usepackage{graphics}
\usepackage{amsmath}
\usepackage{rotating}
\usepackage{lscape}
\usepackage{longtable}

\begin{document}

\title{Comparison of physical processes of atom - surface scattering computed by 
classical and quantum dynamics}
\author{Tapas Sahoo}
\affiliation{S. N. Bose National Centre for Basic Sciences, Salt Lake, Kolkata 700106, India}
\date{\today}

\begin{abstract}
We have performed classical and quantum dynamical simulations to calculate 
dynamical quantities for physical processes of atom - surface scattering, e.g., 
trapping probability and average energy loss, final angular distribution of 
a particle scattered from a corrugated thermal surface. Here we have restricted 
ourselves to in-plane scattering so that only two degrees of freedom of the 
particle have to be considered - the vertical distance $z$ and the horizontal 
coordinate $x$. Moreover, we assumed further that only the vertical coordinate 
fluctuates due to interaction with thermal phonon bath of the surface. Initial 
phase - space variables of the system and the bath for our classical simulations 
were generated according to Wigner distribution functions which were derived from 
initial wavefunctions of our quantum dynamics. At very low incident energy, we 
have found that the quantum mechanical average energy loss of the escaped  
particle from the corrugated as well as thermal surface are smaller than the 
classical ones at a particular surface temperature. It is important to note that 
the rate of escaping probability of the scattered particle obtained by classical 
simulation increases with increasing surface temperature. On the other hand,
quantum rate is almost temperature independent at 2 meV incident energy of 
the particle, whereas it shows same trend with the classical results at 5 meV 
and the quantum rate is lower than the classical rate. We have also noticed 
that the final angular distributions of the scattered particle both for classical 
as well as quantum dynamics are qualitatively different but the quantities are
more or less temperature independent. 

\end{abstract}

\maketitle

\renewcommand{\theequation}{1.\arabic{equation}} \setcounter{section}{0} %
\setcounter{equation}{0}

\section{Introduction}

During past few decades, the trapping as well the effect of quantum diffraction 
on the final angular distribution of an atom scattered on thermal corrugated surface 
have created enormous interests as a paradigm of atom - surface scattering dynamics 
both for experimental\cite{zp-61-95-1930,ssr-4-1-1984,jcp-94-806-1991,jcp-97-4453-1992,
ssr-17-1-1993,cpl-163-111-1989,jcp-91-1942-1989,jcp-90-3800-1989,jcp-122-244713-2005,
epjd-38-129-2006} and theoretical studies\cite{jcp-80-5827-1984,sc-111-461-1981,sc-226-180-1990,jcp-94-1516-1991,jcp-92-680-1990,
prb-79-045424-2009,jcp-130-064703-2009}. As for example, Mullins \textit{et al.}%
\cite{cpl-163-111-1989} reported that the trapping probability of Ar scattered on 
Pt(111) decreases with increasing incident kinetic energy and it follows 
$E_i\cos^n(\theta_i)$ law, where $E_i$ and $\theta_i$ are the incident kinetic 
energy of the particle and the incidence angle relative to the surface normal.
They found that the optimal values of the exponent $n$ are 1.5, 1.0 and 0.5 for 
$T_s$=80, 190, 273K, respectively. On the other hand, in 1930 Estermann and 
Stern\cite{zp-61-95-1930} recorded, experimentally, diffraction of He atom on 
LiF(100) crystal and shown that atom diffraction will be feasible when the 
de Broglie wavelength, $\lambda$, of the impinging atoms is on the order of 
the interatomic spacing of the solid material. Kondo \textit{et al.}\cite%
{jcp-122-244713-2005,epjd-38-129-2006} experimentally measured the angular 
intensity distributions of Ar, CO and N$_{2}$ scattered from a LiF(001) 
surface as functions of surface temperature, incident translational energy, 
and incident azimuthal direction affecting surface corrugation.

Hubbard and Miller \cite{jcp-80-5827-1984} have applied a semiclassical
perturbation (SCP) approximation to calculate the sticking probability for
the He-W(110) and Ne-W(110) systems and they have shown that it works well 
at low surface temperature. Tully have derived explicit expression 
\cite{jcp-92-680-1990} for the sticking probability using the washboard model 
and found that the trapping probability is close to unity at low incidence 
energies and goes to zero as the energy is increased. Fan and Manson 
\cite{prb-79-045424-2009, jcp-130-064703-2009} have developed a classical 
theory for the trapping and desorption of heavy atoms colliding with surfaces which
accounts for both direct scattering and trapping-desorption of the incident
beam. Tully \cite{sc-111-461-1981,sc-226-180-1990} and Head-Gordon \textit{et al.} 
\cite{jcp-94-1516-1991} have performed molecular dynamics simulations for 
the scattering of Ar and Xe on a Pt(111) surface and the Ar-Pt(111) system,
respectively.

Pollak \textit{et al.} have developed a classical\cite{jcp-129-054107-2008,
jcp-130-194710-2009,jcp-132-049901-2010,prb-80-115404-2009,prb-81-049903-2010,
prb-80-165420-2009,prb-81-039902-2010,cp-375-337-2010} as well as a semiclassical 
perturbation\cite{jcp-137-201103-2012,jcp-136-204707-2012,jcp-142-174102-2015}
theories for the scattering of an atom from a metal surface in the limit of weak 
corrugation and assuming weak (linear) coupling of the projectile to the harmonic 
surface phonon modes. More recently they extended the classical perturbation 
theories to include second order corrections to the angular distribution\cite{jcp-140-024709-2014} 
and the sticking probability\cite{jcp-143-064706-2015} of the scattered particle.
In addition, Pollak \textit{et al.}\cite{jpcc-119-14532-2015} developed second 
order semiclassical perturbation theory to the elastic scattering of an atom 
from a corrugated surface to account the asymmetry in the diffraction pattern.
Asaf azuri and Eli Pollak\cite{jcp-143-014705-2015} have performed classical 
and quantum dynamical simulations for the scattering of Ar atom from a frozen 
LiF(100) surface based on a first principles interaction potential.

In this paper, we have carried out both classical and quantum dynamics 
for the model Hamiltonian, where the particle interacts linearly with 
harmonic surface phonon modes and the interaction potential along vertical
direction is Morse. For simplicity, we have considered only 
two degrees of freedom for the particle - one is vertical coordinate $z$
and the other is horizontal coordinate $x$. Here only the vertical motion, 
which is coupled to an infinite bath of harmonic oscillators, the vertical 
phonon modes. 

The main goal of the present article is to compare the trapping probability,
energy of the escaped particle and the effect of diffraction on the final 
angular distribution computed by classical trajectory and quantum dynamics 
simulations. For quantum dynamics, we have used the Heidelberg Multi-Configuration
Time Dependent Hartree (MCTDH) package\cite{mctdh:package,mey90:73,man92:3199,
bec00:1,mey03:251,mey09:book}. In case of classical simulations, 
the initial phase - space variable are generated by Wigner distribution.
 Moreover, I have performed the dynamical simulations on the model Hamiltonian 
 considering zero incidence angle. 

This article is organized as follows: Section II describes in detail the model 
Hamiltonian, initial wavefunctions for quantum dynamics, initial phase-space 
variable for classical trajectory simulations and the mathematical definitions 
of dynamical quantities. Section III discuses the numerical details of the 
dynamics and compares the classical dynamical quantities with the corresponding 
quantum results elaborately. Finally, some conclusions are gathered in
Section V.

\renewcommand{\theequation}{2.\arabic{equation}} \setcounter{section}{1} %
\setcounter{equation}{0}

\section{Theoretical framework}

\subsection{The model Hamiltonian}

As already mentioned in Sec. I, we restrict our quantum and classical dynamical 
simulations to in-plane scattering so that only two degrees of freedom, e.g., the 
vertical distance $z$ and the horizontal coordinate $x$ of the particle are considered. 
We consider a particle, with mass $M$ and incident negative momentum $p_{z_i}$ and 
positive momentum $p_{x_i}$, which interacts linearly with the (harmonic) surface phonon 
modes. For simplicity, we take into account that only the vertical coordinate of the 
particle interacts with thermal phonon bath of the surface. Assuming weak interaction 
between the projectile and the metal surface, one may consider the model Hamiltonian 
governing the scattering event is
\begin{equation}
\hat{H}=\frac{\hat{p}_{z}^{2}+\hat{p}_{x}^{2}}{2M}+\hat{V}(z,x)+\frac{1}{2}\sum_{j=1}^{N}\left( \hat{p}_{j}^{2}+\omega
_{j}^{2}\left[ \hat{x}_{j}-\frac{c_{j}}{\sqrt{M}\omega _{j}^{2}}\hat{V}^{\prime }(z)%
\right] ^{2}\right),  \label{2.1}
\end{equation}%
where the coordinate ($x_{j}$) and momenta ($p_{j}$) of the $j$-th bath oscillator are 
mass weighted. $\omega _{j}$ and $c_{j}$ are the frequency and the coupling constant 
for the $j$-th bath oscillator, respectively. 

The interaction potential $V(z,x)$ is mainly depends on the instantaneous distance 
$z+\frac{1}{\sqrt{M}}\sum_{j=1}^{N}c_{j}x_{j}$ of the particle from the surface. 
Typically the surface is corrugated so that this distance is modulated periodically
by the corrugation function $h(x)$. In the simplest possible approximation, we consider 
a sinusoidal surface corrugation with period $l$ (the lattice length) and amplitude $h$.
Therefore the form we shall use throughout this article for the interaction potential 
of the atom with the surface in the absence of fluctuations will be
\begin{equation}
\hat{V}(z,x)=\hat{V}(z)+\frac{h\sin(2 \pi \hat{x})}{l}\hat{V}^{\prime}(z).  \label{2.2}
\end{equation}%
$V(z)$ is the potential in the vertical direction, $V^{\prime }(z)$ is the derivative 
with respect to the vertical coordinate. We assume that the corrugation is weak, 
that is $h$ is much smaller than the lattice lengths $l$, or the characteristic 
length scale associated with the vertical potential.

The system potential $V(z)$ vanishes for large vertical distances, it has the qualitative 
form of a Morse potential. The bath Hamiltonian using mass weighted coordinates ($x_{j}$) 
and momenta ($p_{j}$) is 
\begin{equation}
\hat{H}_{B}=\frac{1}{2}\sum_{j=1}^{N}[\hat{p}_{j}^{2}+\omega _{j}^{2}\hat{x}_{j}^{2}].
\label{2.3}
\end{equation}%

\subsection{Initialization}

We have performed quantum dynamical calculations, by employing the Heidelberg MCTDH Package\cite{mctdh:package,mey90:73,man92:3199,bec00:1,mey03:251,mey09:book}, 
on the Hamiltonian (see Eq. \ref{2.1} ) that defines the atom - surface scattering. Initially, 
one can consider the total wavefunction (at time $t$ = 0) as a Hartree product of one-dimensional 
functions:
\begin{eqnarray}
\Psi(z, x, {{\bf x}},t=0) = \varphi(z)\varphi(x)\prod_{j=1}^{N}\varphi(x_{j}). 
\label{2.4}
\end{eqnarray}%
The initial system wavefunction is a two dimensional Gaussian centered around $z_{i}$ and 
$x_{i}$ (= 0) with incident average momentum $p_{z_{i}}$ and $p_{x_{i}}$ (= 0):
\begin{eqnarray}
\psi(z, x, t=0) = \varphi(z)\varphi(x) 
=  \frac{1}{\sqrt{2\pi\Delta z^2\Delta x^2}}\exp \Big[-\Big(\frac{z-z_{i}}{2\Delta z}\Big)^2%
+\frac{ip_{z_{i}}(z-z_{i})}{\hbar}- \Big(\frac{x}{2\Delta x}\Big)^2\Big], 
\label{2.5}
\end{eqnarray}%
where $\Delta z$ and $\Delta x$ are the width of the Gaussian wavefunctions along $z$ and 
$x$ coordinates, respectively. Here we assume that the average initial position of the 
incident particle is 80 $\AA$ which is far away from the metal surface so that the particle
is not coupled to the phonon modes. On the other hand, one can easily calculate the average 
initial momentum ($p_{z_{i}}$) by incident energy of the particle. In addition we take 
$\Delta z$ = 5 a.u. but $\Delta x$ = 40 a.u., which is larger than the lattice length 
(3.61 $\AA$), to get diffraction pattern in case of quantum dynamical computations. 
The Wigner distribution function for classical trajectory simulations to generate initial 
positions and momenta of the incident atom along $z$ and $x$ coordinates is
\begin{eqnarray}
\rho_{S,W} = \frac{1}{\pi^2\hbar^2}\exp \Big[-\frac{1}{2}\Big(\frac{z-z_{i}}{\Delta z}\Big)^2 
- \frac{1}{2}\Big(\frac{x}{\Delta x}\Big)^2
-\frac{2\Delta z^2 (p_{z}-p_{z_{i}})^2}{\hbar^2}
-\frac{2\Delta x^2 p_{x}^2}{\hbar^2} \Big]. 
\label{2.6}
\end{eqnarray}%

The initial wavefunction for the $j$-th bath mode (using mass weighted coordinates) 
is also taken as an eigenfunction of a harmonic oscillator for a particular quantum 
state. Initially the incident particle is sufficiently far away from the surface such 
that the interactions between the particle and the surface phonons are vanished. As a 
result the initial conditions for the bath phase space variables for classical simulations 
are taken from the following Wigner representation of the thermal bath: 
\begin{eqnarray}
\rho_{B,W}({\bf p,x}) & \equiv & \Big \langle \frac{\exp(-\beta H_B)}{Z_B} \Big \rangle_W \notag \\
&=& \prod_{j=1}^{N} \Big( \frac{\nu_j}{\pi \hbar} \exp \Big(-\frac{\nu_{j}}{\hbar \omega_{j}}
(p_{j}^2+\omega_{j}^2 x_{j}^2)\Big) \Big),
\label{2.7}
\end{eqnarray}%
with 
\begin{equation}
\nu_{j}\equiv \tanh\Big(\frac{\hbar \beta \omega_{j}}{2}\Big)
\label{2.8}
\end{equation}%
and $\beta =1/(k_{B}T)$.
 
\subsection{Dynamical quantities}
In our classical trajectory calculations, the escaping probability is defined as:
\begin{eqnarray}
P_{escape}(t)=\frac{{\displaystyle \sum_{k=1}^{N}}\Theta(z_{k}(t)\geq z_{0})}{N},
\label{2.9}
\end{eqnarray}
where $\Theta(z_{k}\geq z_{0})$ is the Heaviside function. It is zero if the 
distance between the particle and the surface ($z_{k}$) of $k$-th trajectory 
is less than $z_{0}$ = 50 $\AA$, otherwise, it is one. $N$ is total number of 
trajectories. Now one can easily determine the trapping probability $P_{trap}$
of the particle scattered from surface by subtracting the escaping probability 
from unity, i.e.
\begin{eqnarray}
P_{trap}(t)=1-P_{escape}(t).
\label{2.10}
\end{eqnarray}
 One may also define the average energy of the escaped particle as:
\begin{eqnarray}
\langle E_{es}(t)\rangle	=\frac{{\displaystyle \sum_{k=1}^{N}}E_{k}(t)\Theta(z_{k}(t)\geq z_{0})}{N},
\label{2.11}
\end{eqnarray}
where, $E_{k}(t)$ is the total energy of the particle for $k$-th trajectory. 

Similarly, the trapping probability and the average energy of the escaped particle 
computed by quantum dynamics are defined as:
\begin{eqnarray}
P_{trap}(t)	=1-\frac{\langle\Psi(z,x,\{q_{j}\}, t)|\Theta(z\geq z_{0})
|\Psi(z,x,\{q_{j}\}, t)\rangle}{\langle\Psi(z,x,\{q_{j}\}, t)|\Psi(z,x,\{q_{j}\},t)\rangle},
\label{2.12}
\end{eqnarray}
\begin{eqnarray}
\langle E_{es}(t)\rangle	=\frac{\langle\Psi(z,x,\{q_{j}\}, t)|\hat{H}_{z,x}\Theta(z\geq z_{0})
|\Psi(z,x,\{q_{j}\}, t)\rangle}{\langle\Psi(z,x,\{q_{j}\}, t)|\Psi(z,x,\{q_{j}\}, t)\rangle},
\label{2.13}
\end{eqnarray}
where $\Psi(z,x,\{q_{j}\})$ and $\hat{H}_{z,x}$ are total wavefunction of the 
atom-surface interaction and the hamiltonian operator of the particle, respectively. 

The normalized energy of the scattered particle is calculated by the following 
expressions:
\begin{eqnarray}
\langle E({p_x})\rangle = \dfrac{\int dp_{z} \dfrac{p_{z}^{2}+p_{x}^{2}}{2M}P(p_{z},p_{x})}
{\int dp_{z} P(p_{z},p_{x})},
\label{2.15}
\end{eqnarray}
$P(p_{z},p_{x})$ is two dimensional density along $x$ and $z$ degrees of freedom. 
Finally, one can easily transformed $\langle E({p_x})\rangle$ to the normalized 
energy of the particle as a function of diffraction quantum number, $n$. $n$ is 
defined through the change of the horizontal component of the wavevector 
$\Delta k_{x}=k_{x_{f}}-k_{x_{i}}$ and guided by Bragg's law as,
\begin{eqnarray}
n = \frac{l\Delta k_{x}}{2\pi}=\frac{l\Delta p_{x}}{2\pi\hbar}.
\label{2.16}
\end{eqnarray}

To calculate the above quantum quantities at a particular surface temperature, 
we have performed quantum dynamics for many number of configurations of the bath 
and then taken average of the dynamical quantities over the total number of 
configurations as following: 
\begin{eqnarray}
\langle A \rangle  = \frac{\sum_{k}A\exp(-E_{k}/k_{B}T)}{\sum_{k}\exp(-E_{k}/k_{B}T)},  \label{2.17} 
\end{eqnarray}
where the index $k$ is generated by the different quantum states of harmonic oscillators.
In our quantum calculations, we have taken total 1296 configurations to compute the 
above quantities upto 80K. Moreover, it is important to note that the maximum allowed 
quantum numbers are 8, 5, 3, 2, 1, 0, 0, 0 for 8 harmonic oscillators of lower frequency 
to higher one, respectively.

\renewcommand{\theequation}{3.\arabic{equation}} \setcounter{section}{2} %
\setcounter{equation}{0}

\section{Numerical results}

A model He-solid surface system has been considered to investigate the physical 
processes of scattering at very low incident energies (2 and 5 meV) of the 
particle and at low temperatures. In this simulations, we have considered 
Morse potential along vertical direction,
\begin{eqnarray}
V(z)=V_{0}(1-\exp(-\alpha z))^2-V_{0},
\end{eqnarray}
where $V_{0}$ and $\alpha$ are a physisorption well depth and a stiffness 
parameter, respectively. The parameters used in this study are $V_0$=34.85 meV, 
$\alpha$=0.5 $ \mathring{A}^{-1}$. In addition, the reduced friction coefficient, 
$\tilde{\gamma}=0.005$ and the corrugation height, $h$ is 0.1 a.u. Here we have 
taken mass of the particle as a mass of Helium atom, $M=4.002602 $ a.m.u. to show 
quantum effect for the atom - surface event. On the other hand, other parameters 
have been chosen arbitrarily so that one can easily compare the classical and 
quantum dynamical results.

\begin{figure}[tph]
\centering
\begin{tabular}{r@{\extracolsep{0pt}.}lc}
\includegraphics[width=10cm,height=10cm,keepaspectratio]{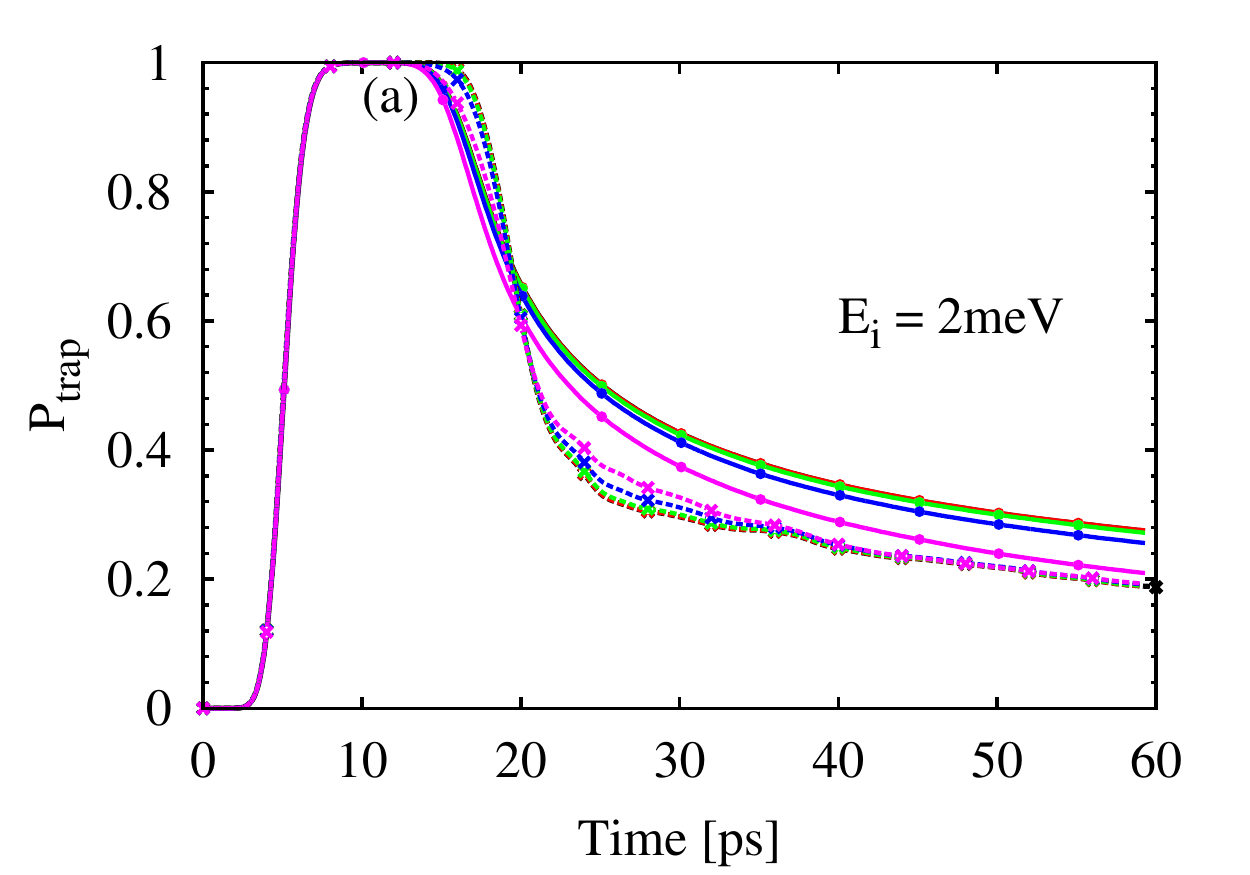}\tabularnewline
\includegraphics[width=10cm,height=10cm,keepaspectratio]{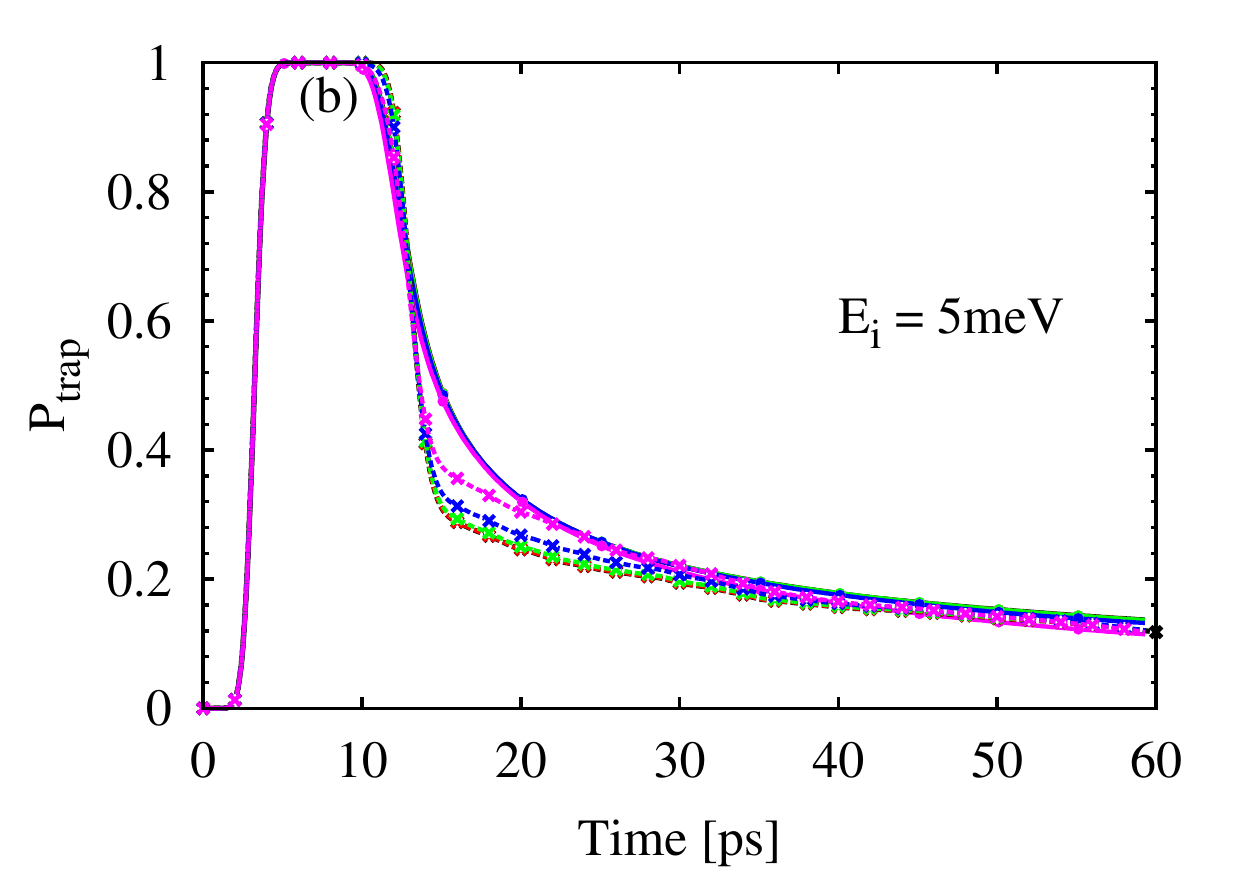}\tabularnewline
\end{tabular}
\caption{(color online) The trapping probability as a function of time at two 
different incident energies, 2 (panel a) and 5 meV (panel b). Black, red, green, 
blue and magenta lines indicate that the quantities are computed at 0, 10, 20, 
40 and 80 K, respectively. In addition, (solid line with circle point) and 
(dashed lines with cross point) are presented by the results obtained from 
classical (CL) and quantum dynamics (QM) simulations, respectively.}
\label{fig:fig1}
\end{figure}

\begin{figure}[tph]
\centering
\begin{tabular}{r@{\extracolsep{0pt}.}lc}
\includegraphics[width=10cm,height=10cm,keepaspectratio]{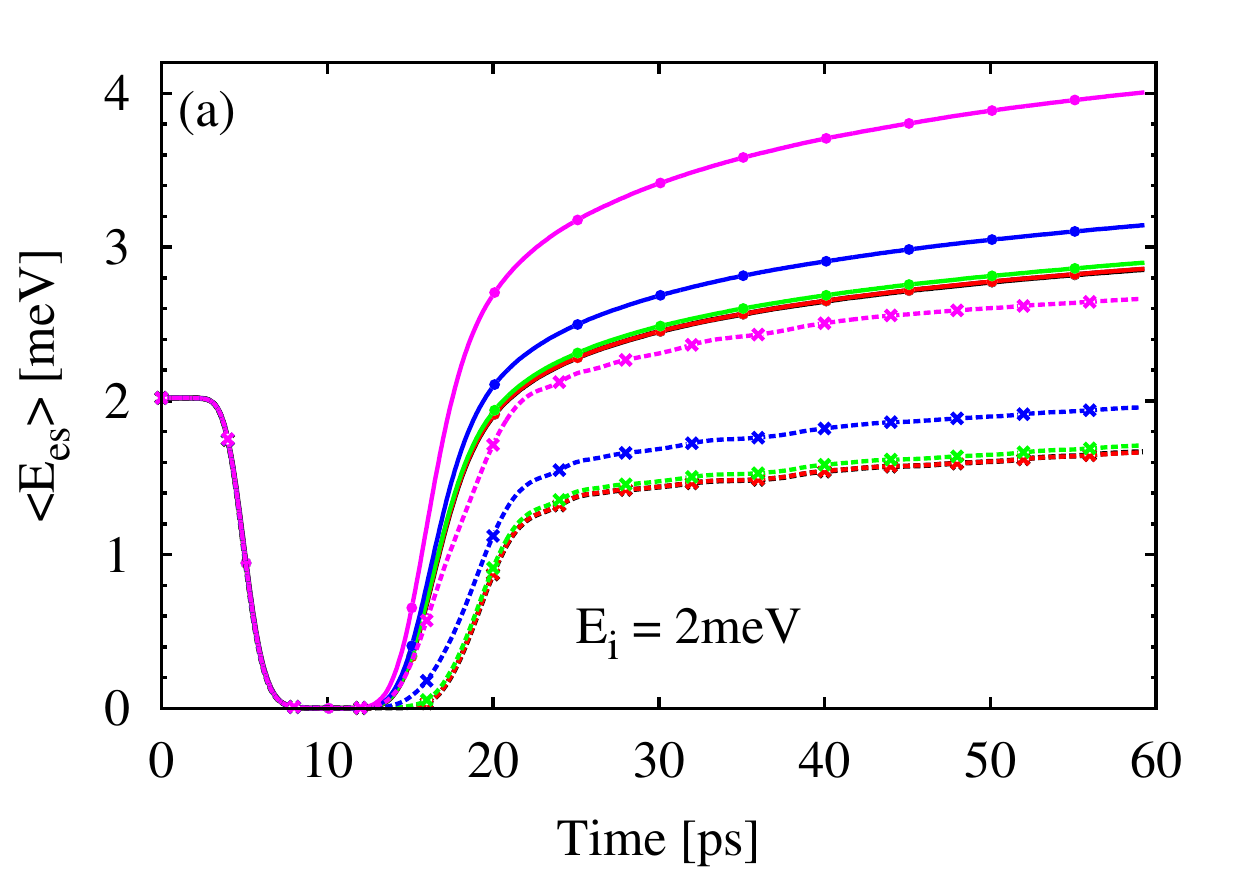}\tabularnewline
\includegraphics[width=10cm,height=10cm,keepaspectratio]{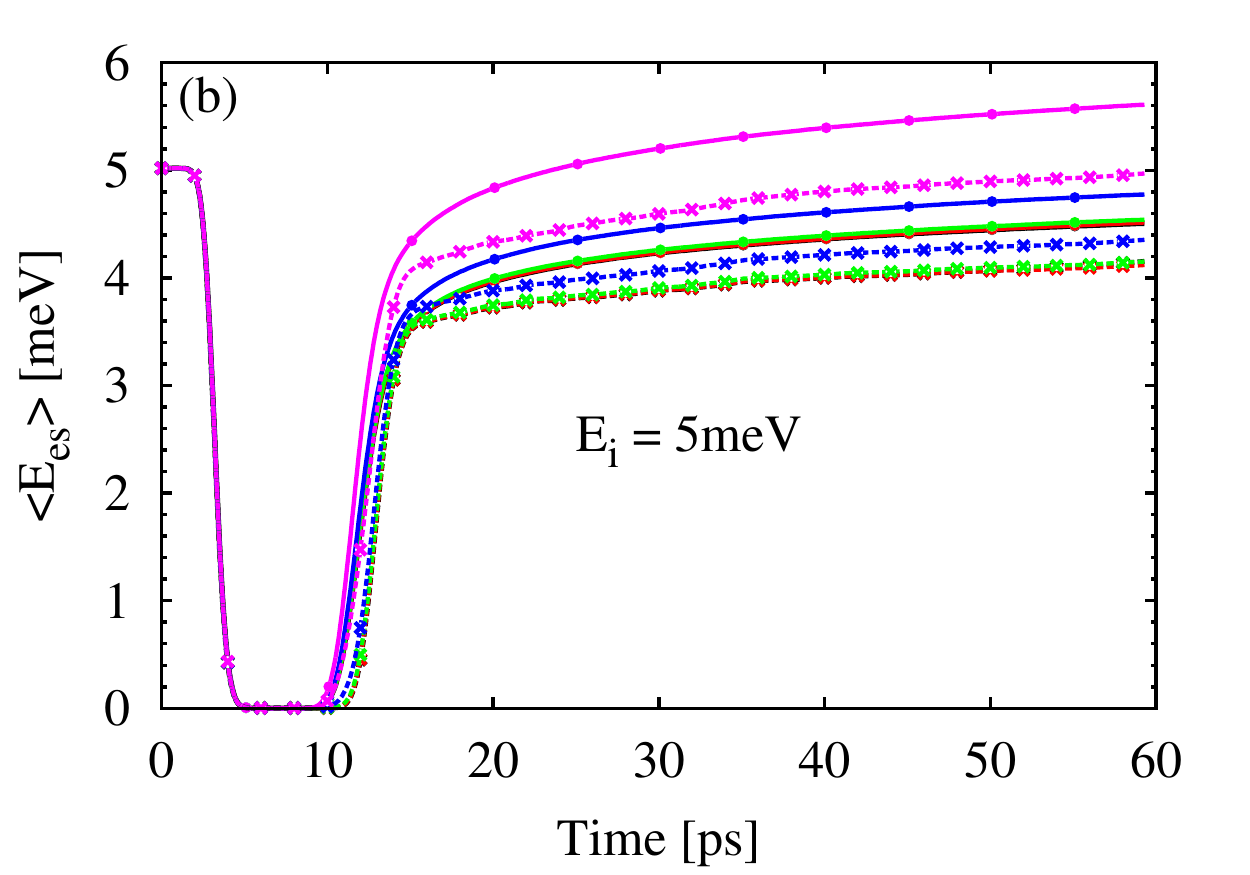}\tabularnewline
\end{tabular}
\caption{(color online) The average energy of the escaped particle as a function 
of time at two different incident energies, 2 and 5 meV. The notation is the same 
as in Fig. \protect\ref{fig:fig1}.}
\label{fig:fig2}
\end{figure}

Our prime goal of this paper is to compare the effect of corrugation as well as 
surface temperature on the dynamical quantities for the atom - surface scattering 
computed by both classical and quantum dynamics. For this purpose we have computed 
the numerically exact solutions of the equations of motion, using a discretized 
bath. The coupling coefficients and frequencies of the discretized bath were 
determined by employing the algorithm of Ref. \cite{jcp-112-47-2000}:
\begin{eqnarray}
\omega_j & = & -\omega_c \ln\Bigg(1-\frac{j}{N+1}\Bigg),  \label{3.1} \\
c_j & = & \sqrt{\frac{2\gamma\omega_j^2\omega_c}{\pi(N+1)}}.  \label{3.2}
\end{eqnarray}
Here, $\omega_c$ is a cut-off frequency. To get converged results, we have used $N$=8 
bath oscillators and $\omega_c=10\omega_0$ for both the numerical simulations.

\begin{table}
\caption{{\bf Data for the rate of escaping of the scattered particle and 
the corresponding errors obtained by fitting the classical trapping probabilities.}}
\begin{tabular}{ |p{3cm}|p{3cm}|p{3cm}|p{3cm}|p{3cm}| }
\hline
\multicolumn{5}{|c|}{Fitted data obtained by classical dynamics} \\
\hline
Temperature & $m$ at 2meV (femtosecond$^{-1}$)& $m$ at 5meV (femtosecond$^{-1}$)& Error at 2meV & Error at 5meV\\
\hline
0K 	& 1.19241e-05 & 1.35107e-05 & 0.522$\%$  & 0.3973 $\%$\\
10K & 1.19258e-05 & 1.37227e-05 & 0.5333$\%$ & 0.438$\%$\\
20K & 1.20895e-05 & 1.37968e-05 & 0.4951$\%$ & 0.4251$\%$\\
40K & 1.31515e-05 & 1.48384e-05 & 0.478 $\%$ & 0.4455$\%$\\
80K & 1.67793e-05 & 1.82815e-05 & 0.4485$\%$ & 0.3346$\%$\\
\hline
\end{tabular}
\label{tab:table1}
\end{table}

\begin{table}
\caption{{\bf Same as Table \ref{tab:table1}, but for quantum dynamical results.}}
\begin{tabular}{ |p{3cm}|p{3cm}|p{3cm}|p{3cm}|p{3cm}| }
\hline
\multicolumn{5}{|c|}{Fitted data obtained by MCTDH} \\
\hline
Temperature & $m$ at 2meV (femtosecond$^{-1}$)& $m$ at 5meV (femtosecond$^{-1}$)& Error at 2meV & Error at 5meV\\
\hline
0K 	& 1.3595e-05  & 1.2926e-05 & 0.899$\%$ & 0.8365$\%$\\
10K & 1.37571e-05 & 1.32683e-05 & 0.9252$\%$ & 0.876$\%$ \\
20K & 1.38088e-05 & 1.34295e-05 & 0.8389$\%$ & 0.8624$\%$\\
40K & 1.37874e-05 & 1.42172e-05 & 0.5905$\%$ & 0.7821$\%$\\
80K & 1.36735e-05 & 1.62797e-05 & 0.4814$\%$ & 0.5878$\%$\\
\hline
\end{tabular}
\label{tab:table2}
\end{table}

\begin{figure}[tph]
\centering
\begin{tabular}{r@{\extracolsep{0pt}.}lc}
\includegraphics[width=10cm,height=10cm,keepaspectratio]{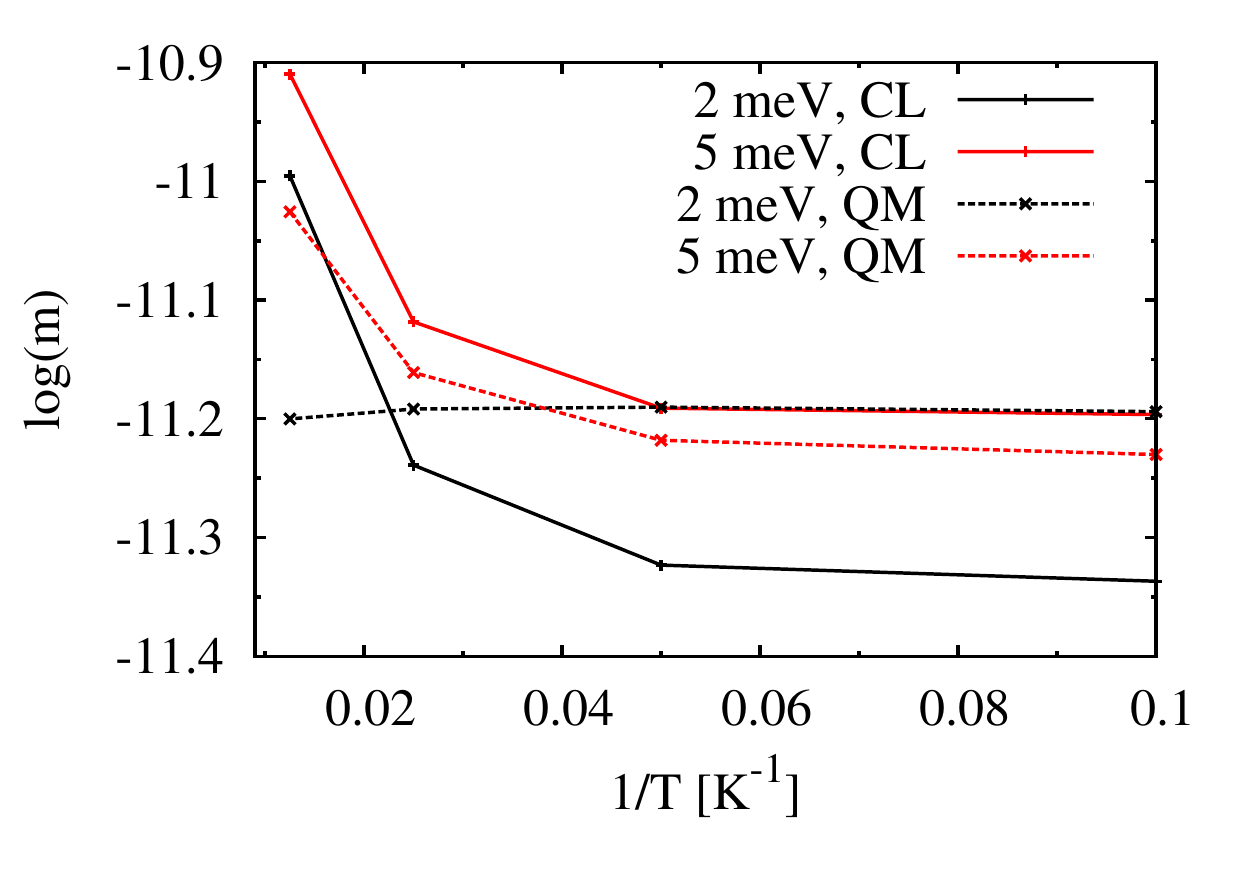}\tabularnewline
\end{tabular}
\caption{(color online) Logarithmic plot of the rate of escaping vs $1/T$, 
where $T$ is the surface temperature. Solid and dashed lines present that
the results are obtained by classical and quantum dynamical simulations, 
respectively. On the other hand, the calculated quantities for the incident 
energies 2 and 5 meV are indicated by the black and red lines, respectively.}
\label{fig:fig3}
\end{figure}
\begin{figure}[htp]
\centering
\begin{tabular}{r@{\extracolsep{0pt}.}lc}
\includegraphics[width=10cm,height=10cm,keepaspectratio]{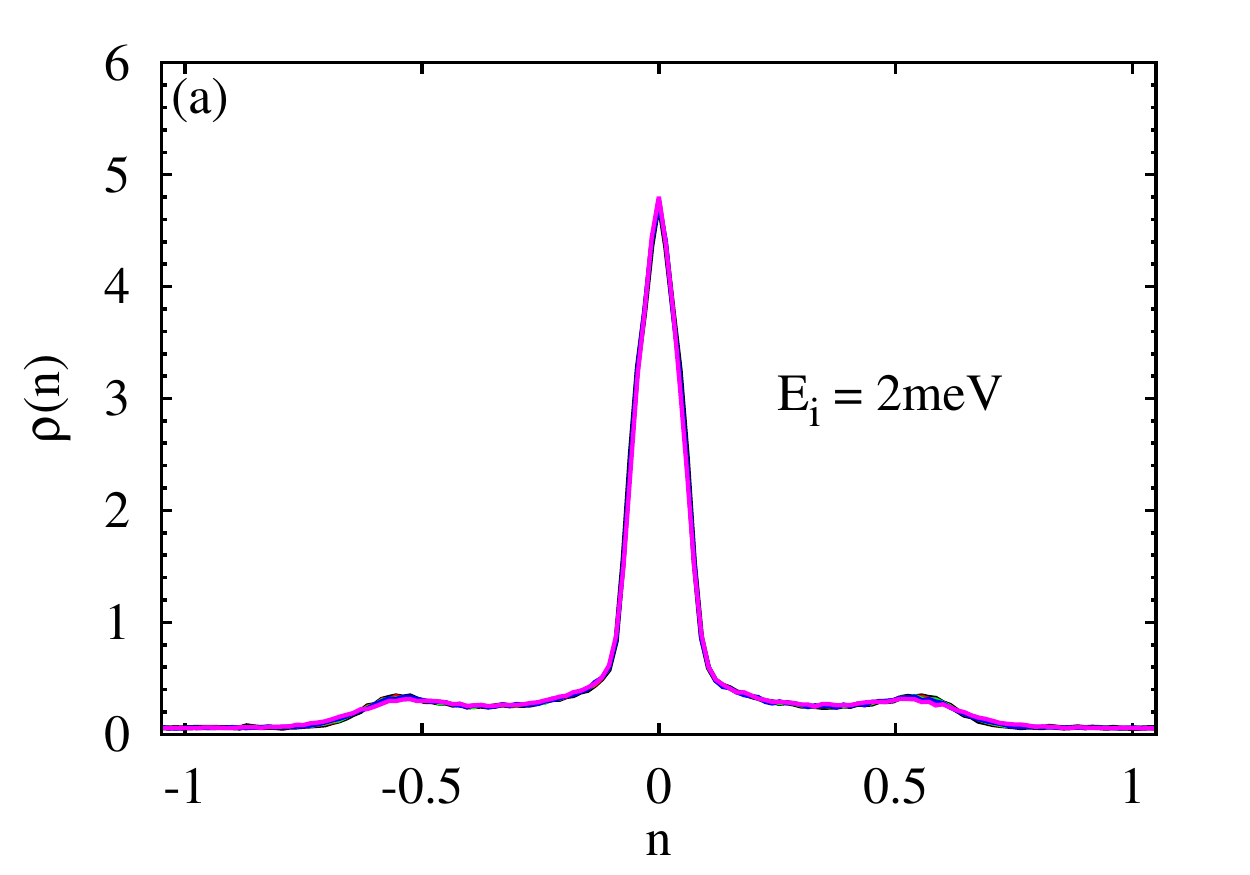}\tabularnewline
\includegraphics[width=10cm,height=10cm,keepaspectratio]{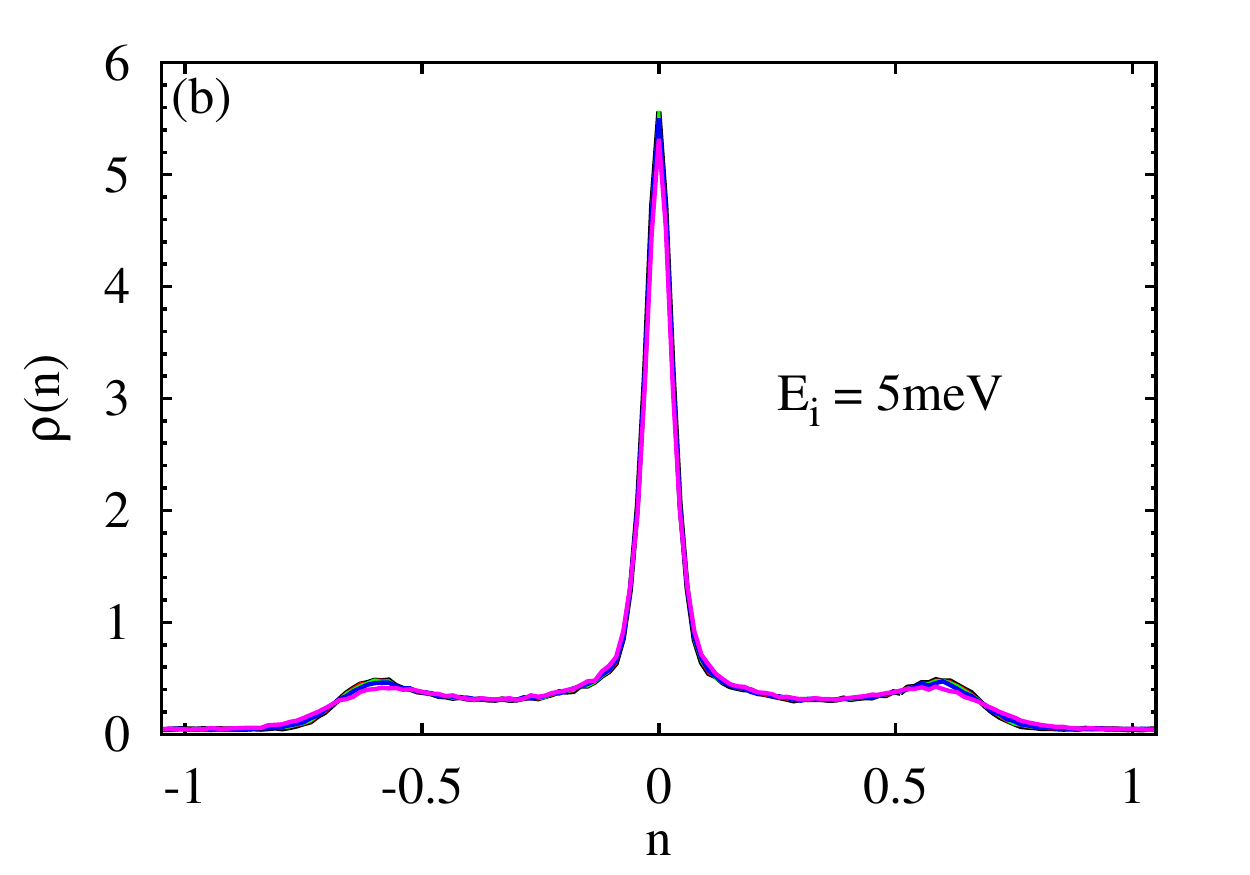}\tabularnewline 
\end{tabular}
\caption{The probability density of the scattered particle, $\rho (n)$, obtained 
by classical trajectory calculation as a function of diffraction quantum number 
($n$) at the incident energies 2 and 5 meV. The notation is the same as in Fig. 
\protect\ref{fig:fig1}.}
\label{fig:fig4}
\end{figure}
\begin{figure}[htp]
\centering
\begin{tabular}{r@{\extracolsep{0pt}.}lc}
\includegraphics[width=10cm,height=10cm,keepaspectratio]{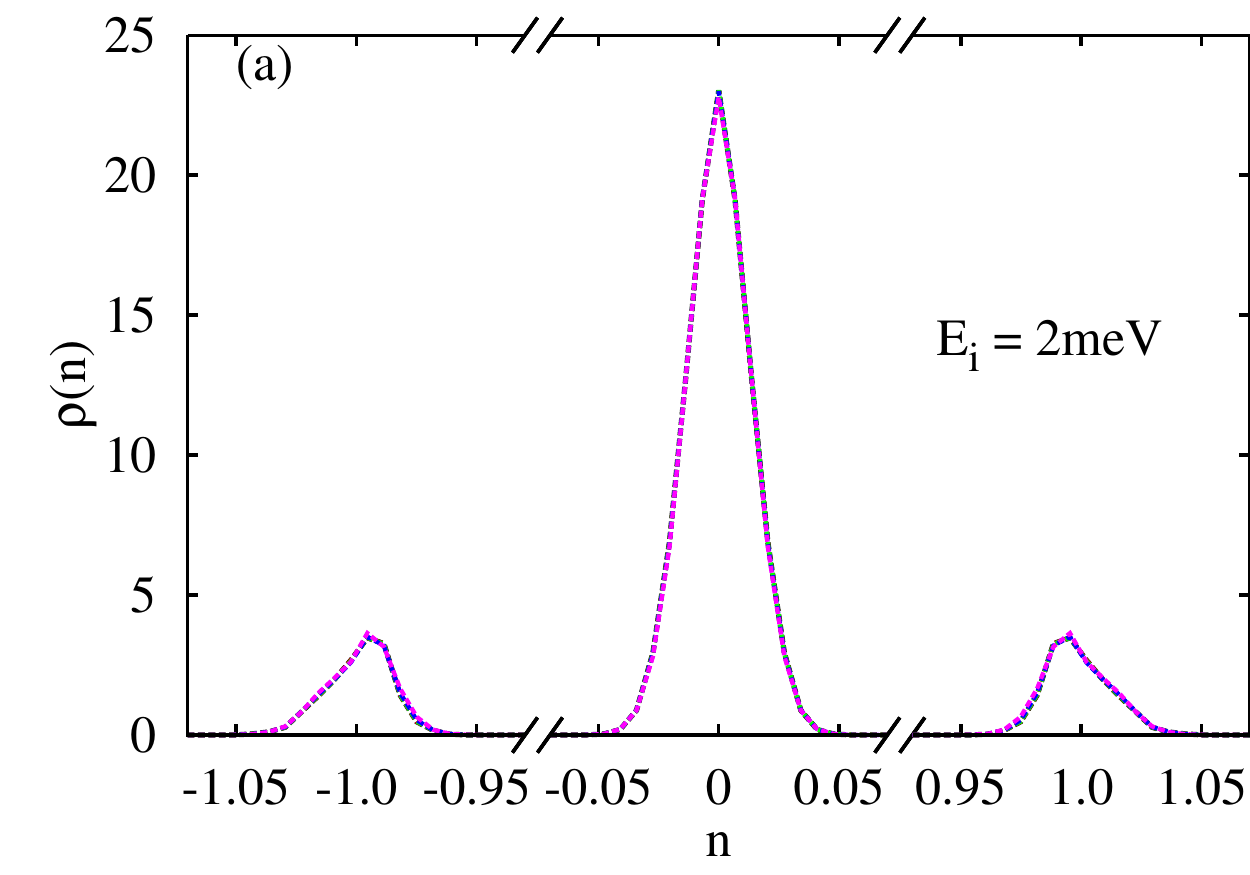}\tabularnewline
\includegraphics[width=10cm,height=10cm,keepaspectratio]{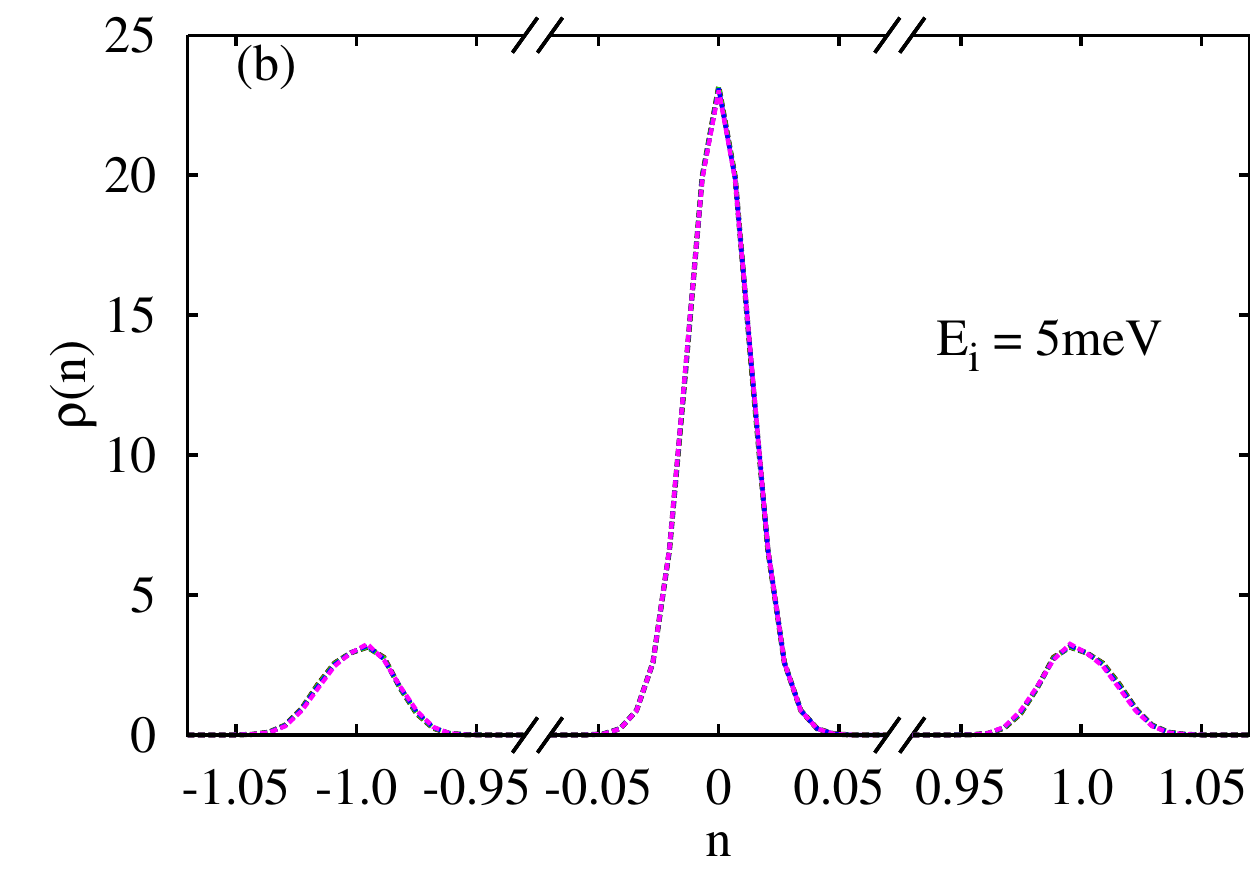}\tabularnewline 
\end{tabular}
\caption{Same as Figure \ref{fig:fig4}, but the quantities are obtained by
performing MCTDH simulations.}
\label{fig:fig5}
\end{figure}

In the classical dynamics, the leap-frog algorithm\cite{leapfrog} has been used 
to propagate trajectories according to Hamilton's equation of motion. On the other hand, 
we employed MCTDH package\cite{mctdh:package,mey90:73,man92:3199,bec00:1,mey03:251,
mey09:book} for the quantum dynamics, where the constant mean-field (CMF) 
integration scheme with a variable step size were used. Within the CMF scheme 
the Runge Kutta (RK8) of order 8 and the short iterative Lanczos (SIL) integrators 
were considered for propagating the single particle functions (spf) and the MCTDH 
coefficients, respectively. The error tolerances for the CMF, RK8 and SIL 
integrators were taken to be $10^{-7}$. In particular, the primitive basis along 
the vertical ($z$) and horizontal ($x$) degree of freedom were described by 
exponential-DVR and Fast Fourier Transform (FFT), respectively. On the other hand, 
the bath ($x_{j}$, $j$=1-8) degrees of freedom were presented by Harmonic Oscillator 
(HO) DVR. We have taken 3071 grid points with the end points at $z_{min}$ = -10 a.u. 
and $z_{zmax}$ = 1200 a.u. and 1536 grid points with the end points at $x_{min}$ 
= -500 a.u. and $x_{max}$ = 500 a.u. along $z$ and $x$ directions, respectively. 
In addition, we have used 15 and 9 spf for $z$ and $x$ coordinates, respectively, 
to get converged results. The HO DVR were taken as 13, 9, 5, 5, 4, 3, 3, 3 grid 
points for 8 bath degree of freedom of lower frequency to higher one, respectively. 
The equilibrium position of each HO DVR is 0.0 and its frequency was determined 
according to Eq. (\ref{3.1}). For 8 bath modes, modes $\{1, 2\}$, $\{3, 4\}$, 
$\{5, 6\}$, and $\{7, 8\}$ were treated as a combined mode assigned with 7, 4, 3 
and 3 single particle functions, respectively. 

\begin{figure}[htp]
\centering
\begin{tabular}{r@{\extracolsep{0pt}.}lc}
\includegraphics[width=13cm,height=13cm,keepaspectratio]{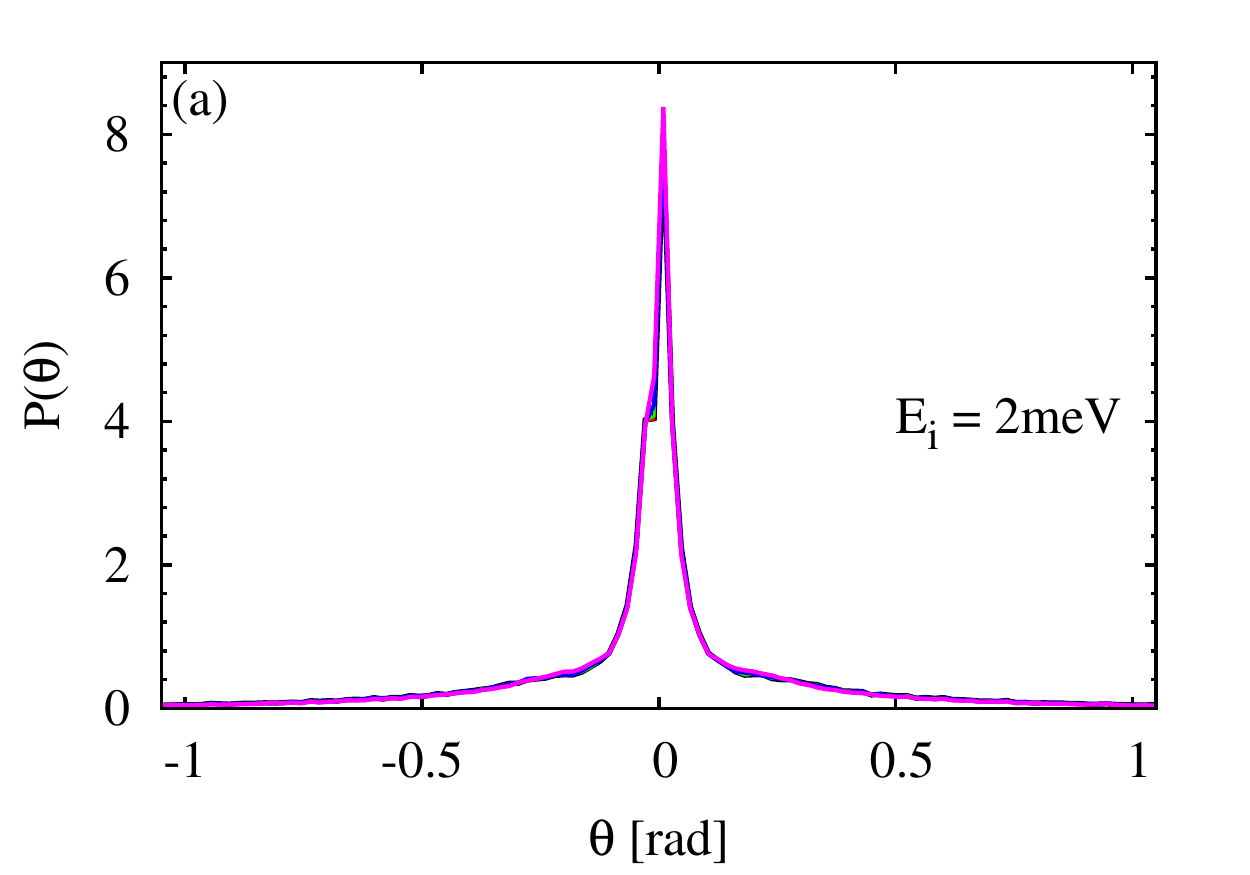}\tabularnewline
\includegraphics[width=13cm,height=13cm,keepaspectratio]{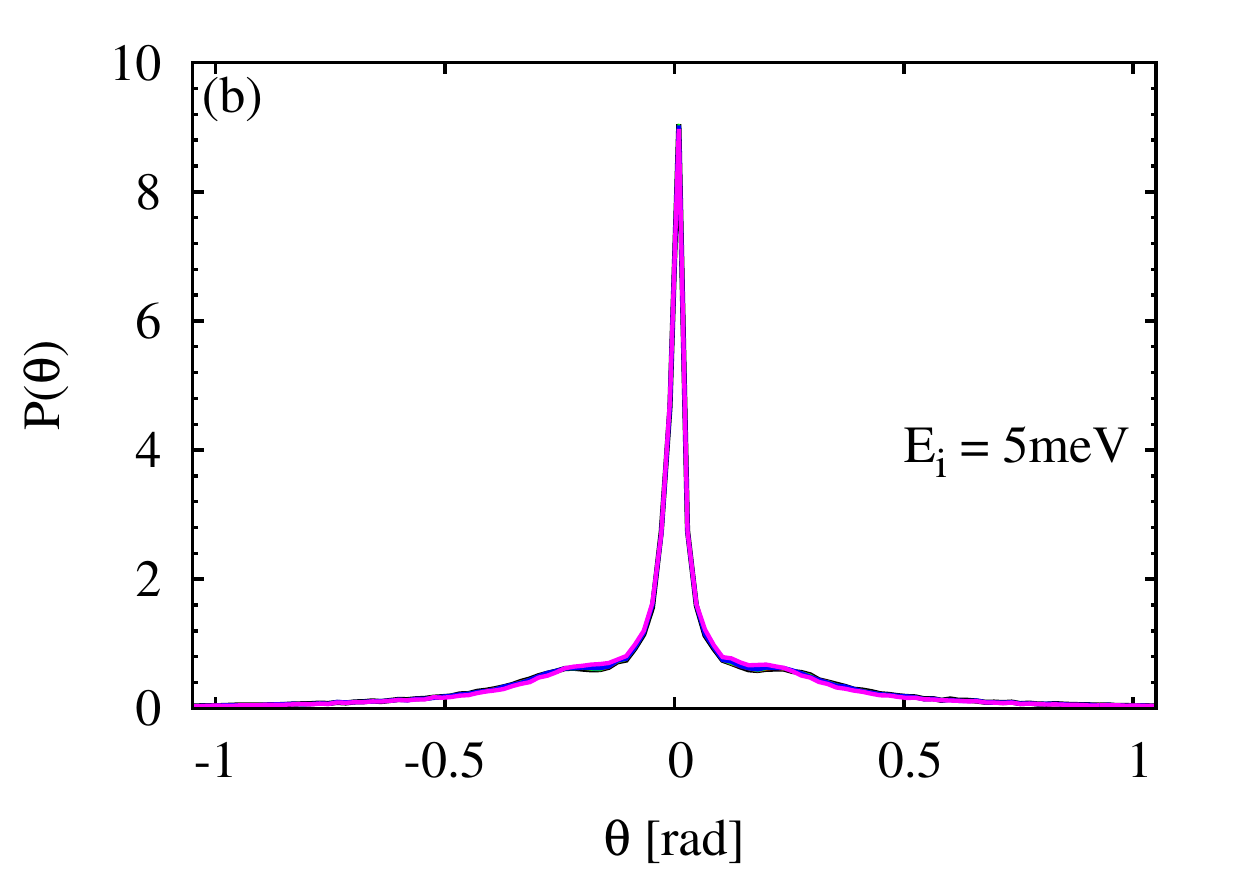}\tabularnewline 
\end{tabular}
\caption{The classical final angular distribution of the scattered particle, $\rho (n)$, as a 
function of diffraction quantum number (n). The notation is the same as in Fig. \protect\ref%
{fig:fig1}. }
\label{fig:fig6}
\end{figure}

\begin{figure}[htp]
\centering
\begin{tabular}{r@{\extracolsep{0pt}.}lc}
\includegraphics[width=13cm,height=13cm,keepaspectratio]{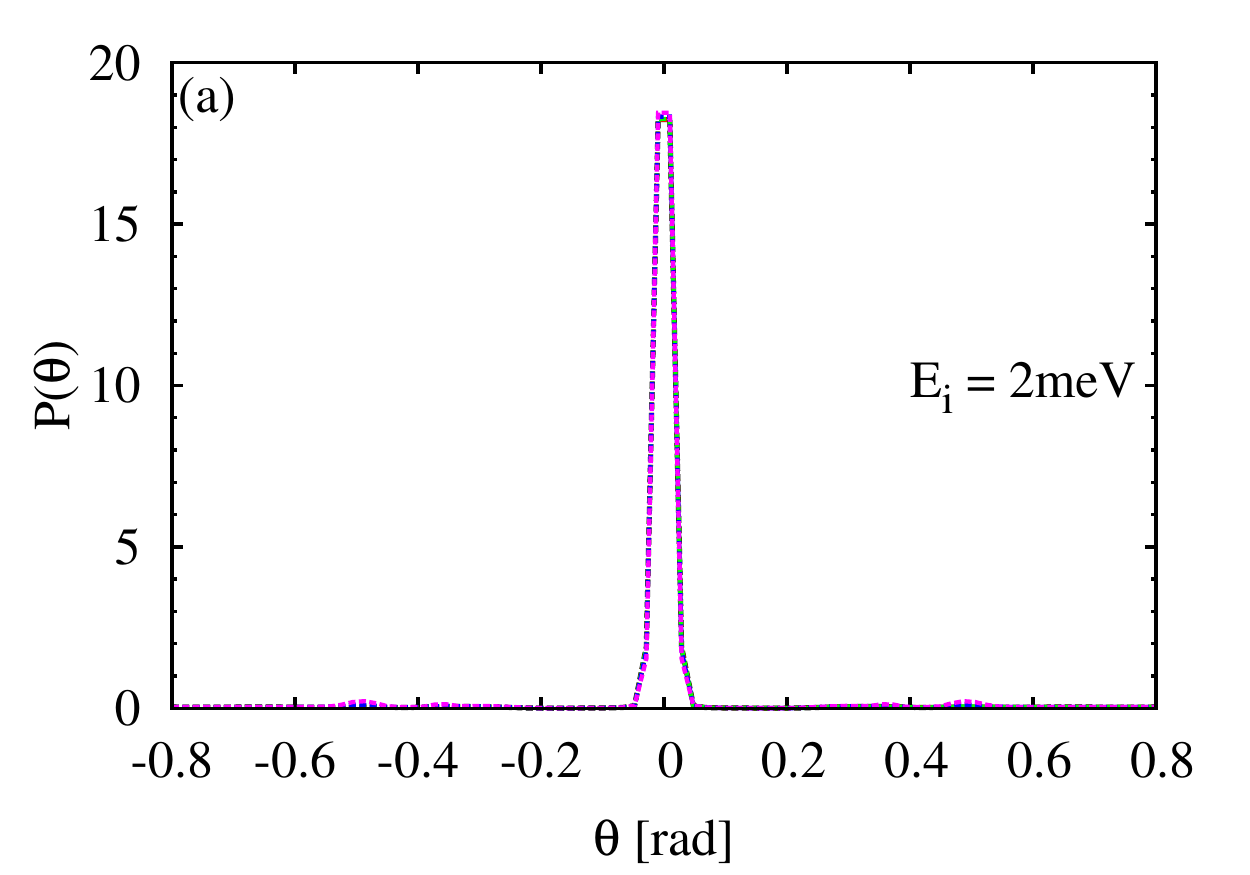}\tabularnewline
\includegraphics[width=13cm,height=13cm,keepaspectratio]{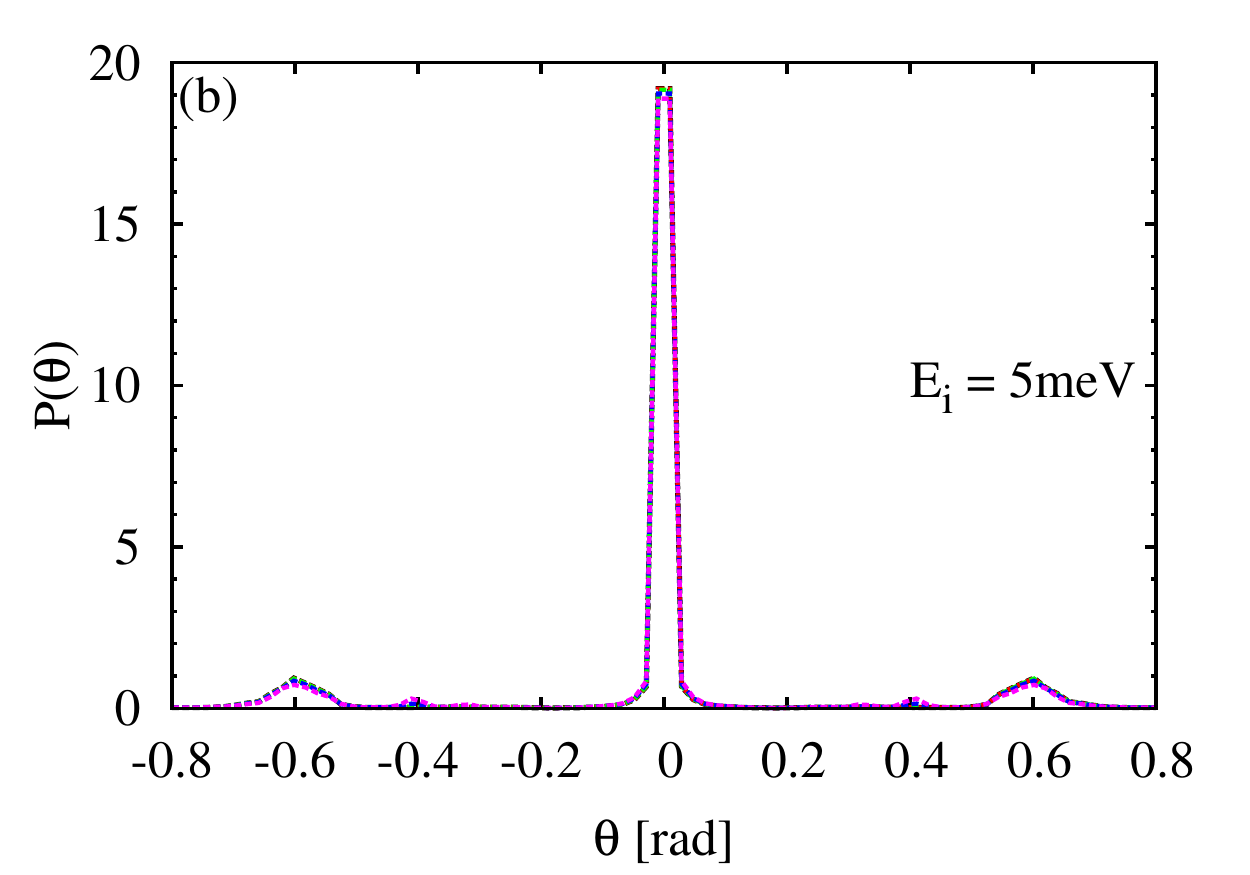}\tabularnewline 
\end{tabular}
\caption{Same as Figure \ref{fig:fig6}, but the quantities are obtained by
performing MCTDH simulations.}
\label{fig:fig7}
\end{figure}

\begin{figure}[htp]
\centering
\begin{tabular}{r@{\extracolsep{0pt}.}lc}
\includegraphics[width=13cm,height=13cm,keepaspectratio]{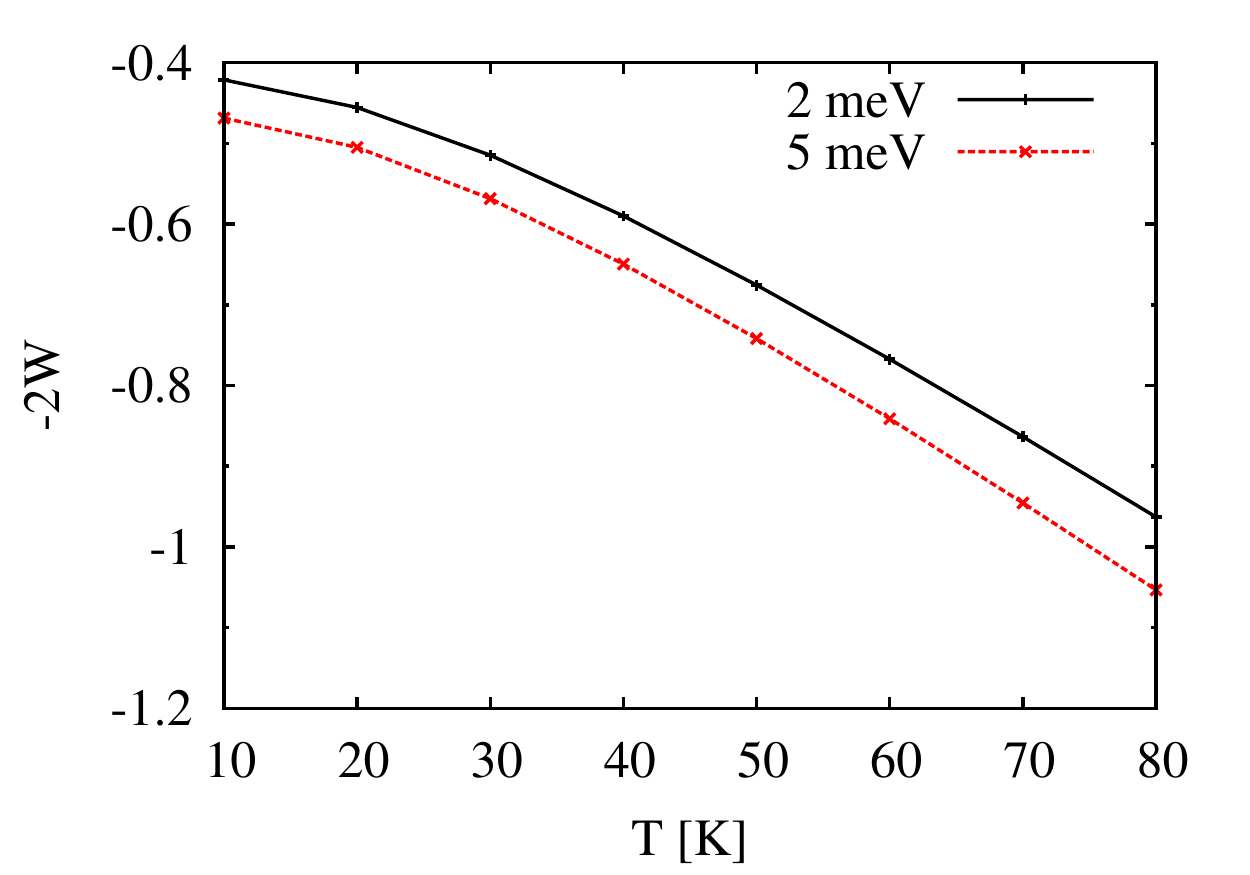}\tabularnewline 
\end{tabular}
\caption{The Debye-Waller factor derived by first order semiclassical perturbation theory
as a function of temperature. Solid black and dashed red lines are presented for two 
incident energies 2 and 5 meV, respectively.}
\label{fig:fig8}
\end{figure}

Figures \ref{fig:fig1} and \ref{fig:fig2} present that the trapping probabilities 
and the average energy of escaped particle as a function of time [picosecond] for 
the incident energies of the particle, 2 (in panel a) and 5 meV (panel b) at 0, 10, 
20, 40 and 80K surface temperatures. Since the initial Gaussian beam is situated 
at 80 a.u., which is far away from the interaction region, the trapping probability 
is zero and the average energy of the escaped particle is same as its incident 
energy at t=0 and then the trapping probability increases and the other one 
decreases when particle approaches to the surface. The trapping and the energy 
of the escaped particle become one and zero, respectively, when the beam is moved 
completely to the interaction region. Finally, the trapping probability 
and the average energy of the escaped particle decreases and increases, 
respectively, as the particle moves away from the interaction region after 
hitting the surface. The trapping probabilities computed by MCTDH dynamics 
at 40 - 60 ps are almost temperature independent. On the other hand, the 
classical trapping probabilities decrease with increasing temperature at 
both the incident energies. Here, we have fitted the numerical data of 
trapping probability at 40 - 60 ps with $c \exp(-m \times time)$ to calculate rate of 
escaping probability, $m$ of the scattered particle from the physisorption 
well. Tables \ref{tab:table1} and \ref{tab:table2} present the fitted data 
of $m$ computed by classical and quantum trapping probabilities at 2 and 5 
meV for various temperatures. At both the energies, the classical rate is 
increased with increasing temperature. On the contrary, the quantum rate is 
temperature independent at 2 meV, while at 5 meV shows the same trend as 
the classical results. In addition, the average energy of the escaped 
particle at asymptotic time increases as the surface temperature increases
and the quantities become positive when the surface energy at a particular 
temperature is greater than the incident energy of the particle, otherwise 
it remains negative in magnitude. Moreover, we have displayed logarithmic 
plot of the rate versus 1/T in Figure \ref{fig:fig3}.
 
In Figures \ref{fig:fig4} and \ref{fig:fig5}, we have plotted the probability 
densities of the scattered particle computed by the classical and quantum dynamics, 
respectively, at final time step 59 picosecond as a function of diffraction quantum 
number $n$. Though the classical probability densities show very small temperature 
dependence, the densities obtained from MCTDH simulations are completely temperature 
independent as the final horizontal momentum of the scattered particle is quantized
by the following:
\begin{eqnarray}
p_{x_{f}}=p_{x_{i}}+\frac{2\pi\hbar}{l}n,
\end{eqnarray}
where $p_{x_{i}}$ and $p_{x_{f}}$ are the initial and final momentum of the particle 
along horizontal direction. In addition, Figures \ref{fig:fig6} and \ref{fig:fig7} 
present the final angular distribution of the scattered particle calculated by 
classical and quantum dynamics, respectively at 59 ps for the incident energies, 
2 (panel a) and 5 meV (panel b). Comparisons of the quantum probability density and 
the final angular distribution with the results obtained from the classical Wigner
computation show that the location of the classical rainbow peaks are not precisely 
mimicked by the quantum probability density or angular distribution. It is important 
to note that, the peaks of the quantum density probabilities as well as the quantum 
angular distributions appeared at diffraction quantum numbers, i.e., -1,0,1. Moreover, 
the profiles of the probability density as well as final angular distribution become 
wider as the incident energy of the incident particle increases. Figure \ref{fig:fig8} 
depicts the Debye-Waller factors, which are computed by semiclassical perturbation (SCP) 
theory, at the two incident energies and according to the scp theory one can expect 
that there is a significant temperature dependence on the quantum angular distributions
(in Figure \ref{fig:fig7}). One should also consider that the SCP formulation is valid
provided that the Eq. (3.5) in our previous article\cite{jcp-143-064706-2015} is satisfied. 
In the present numerical study, the above said condition is not satisfied. That is why 
the comparison between the numerical and the analytical results for final angular 
distribution is not correct.  
 
\begin{figure}[htp]
\centering
\begin{tabular}{r@{\extracolsep{0pt}.}lc}
\includegraphics[width=13cm,height=13cm,keepaspectratio]{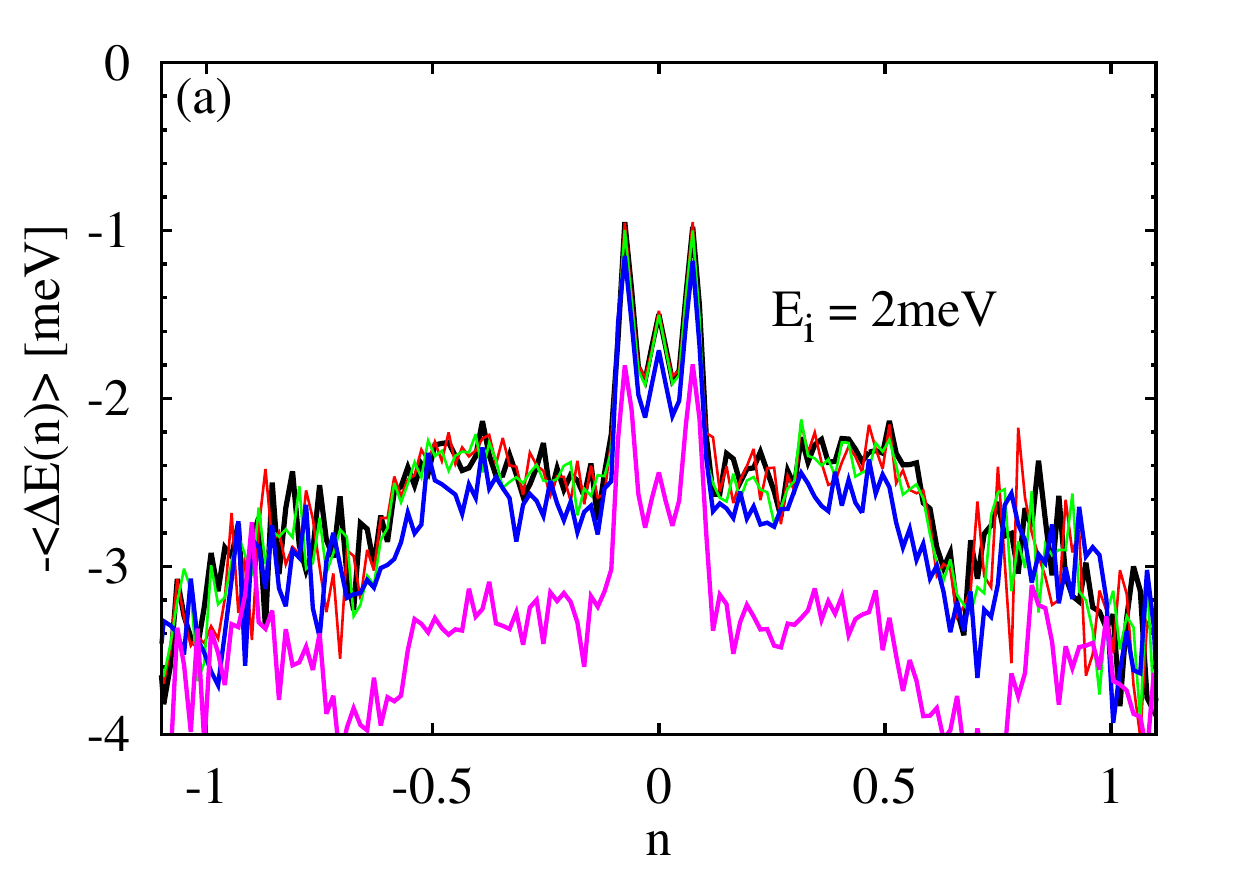}\tabularnewline
\includegraphics[width=13cm,height=13cm,keepaspectratio]{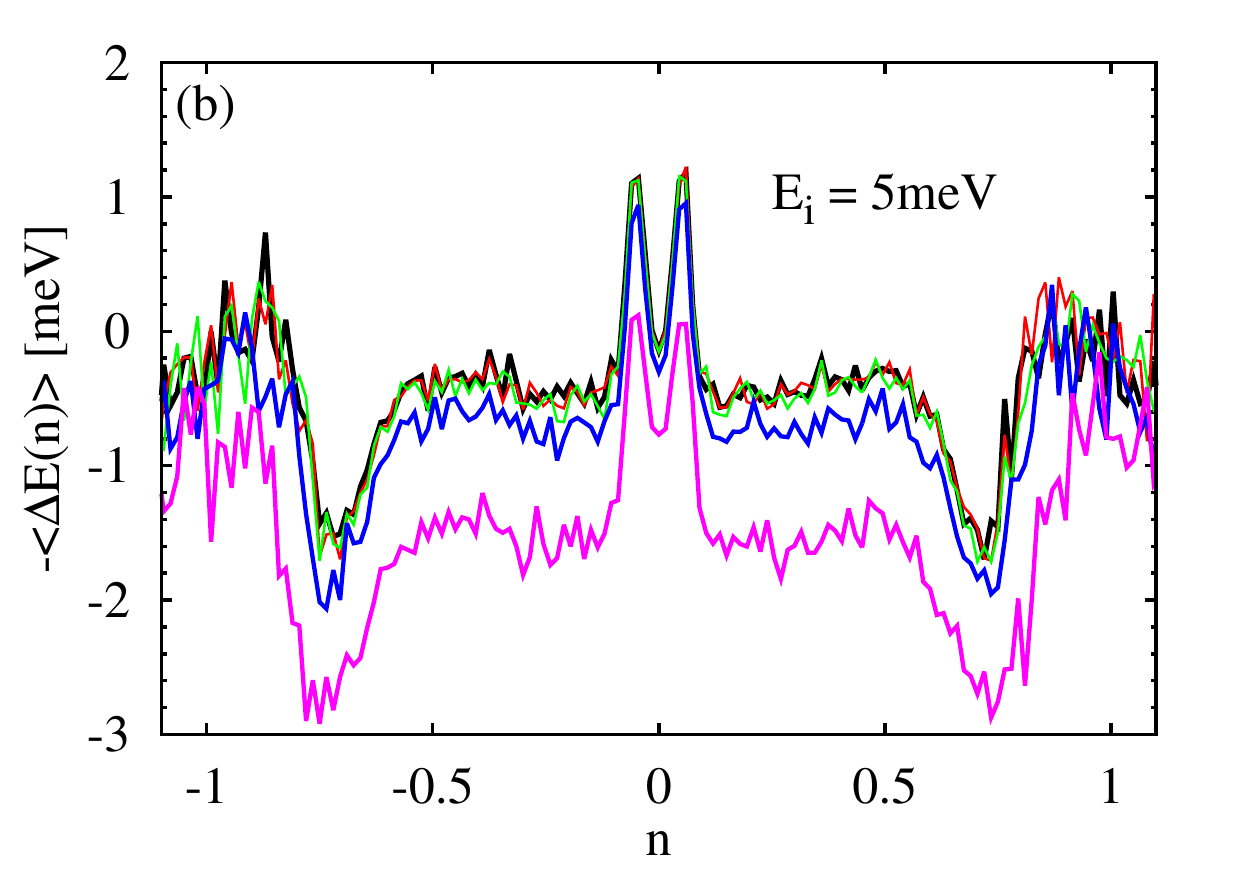}\tabularnewline
\end{tabular}
\caption{The normalized energy loss, $-\langle \Delta E(n) \rangle$, of the particle 
as a function of $n$. The results are obtained by carrying out classical dynamics.
The notation is the same as in Fig. \protect\ref{fig:fig1}.}
\label{fig:fig9}
\end{figure}

\begin{figure}[htp]
\centering
\begin{tabular}{r@{\extracolsep{0pt}.}lc}
\includegraphics[width=13cm,height=13cm,keepaspectratio]{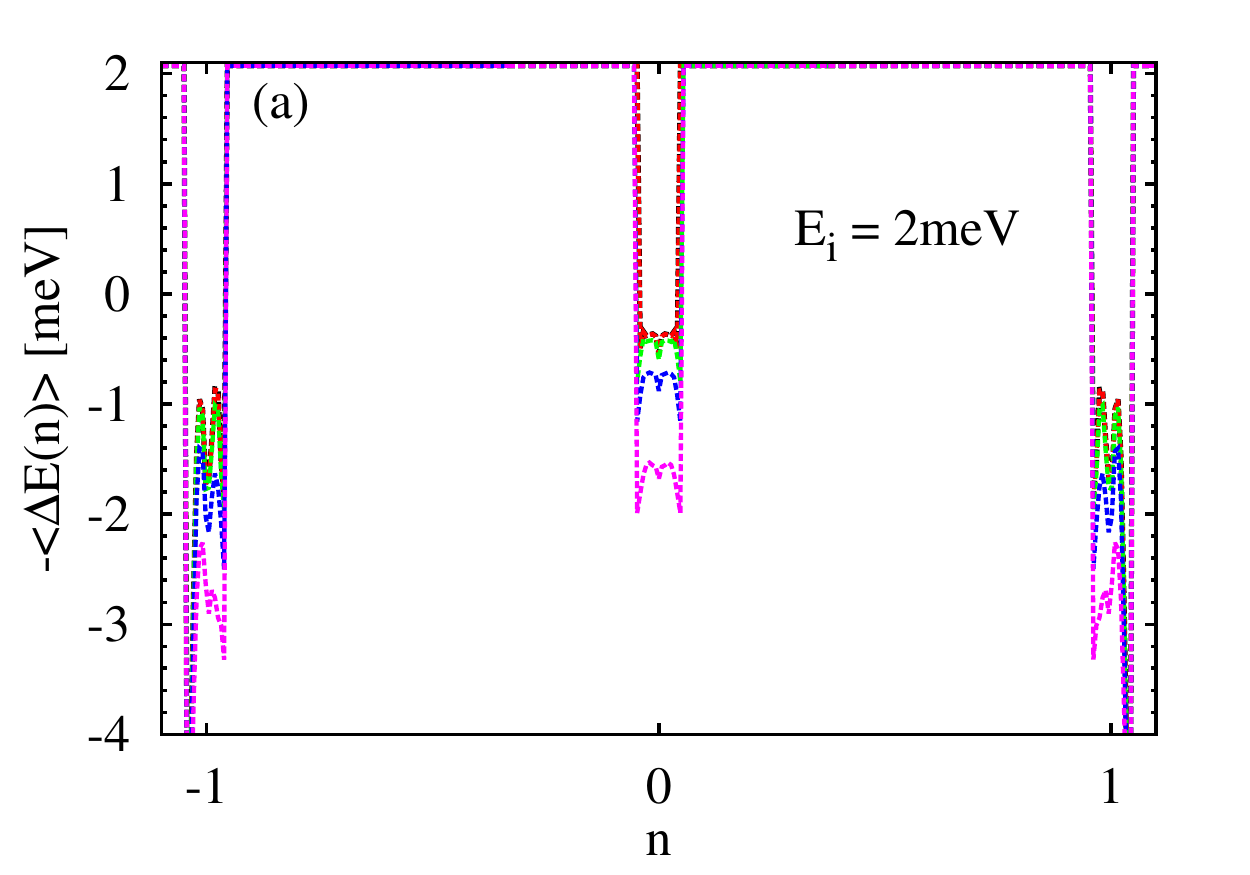}\tabularnewline
\includegraphics[width=13cm,height=13cm,keepaspectratio]{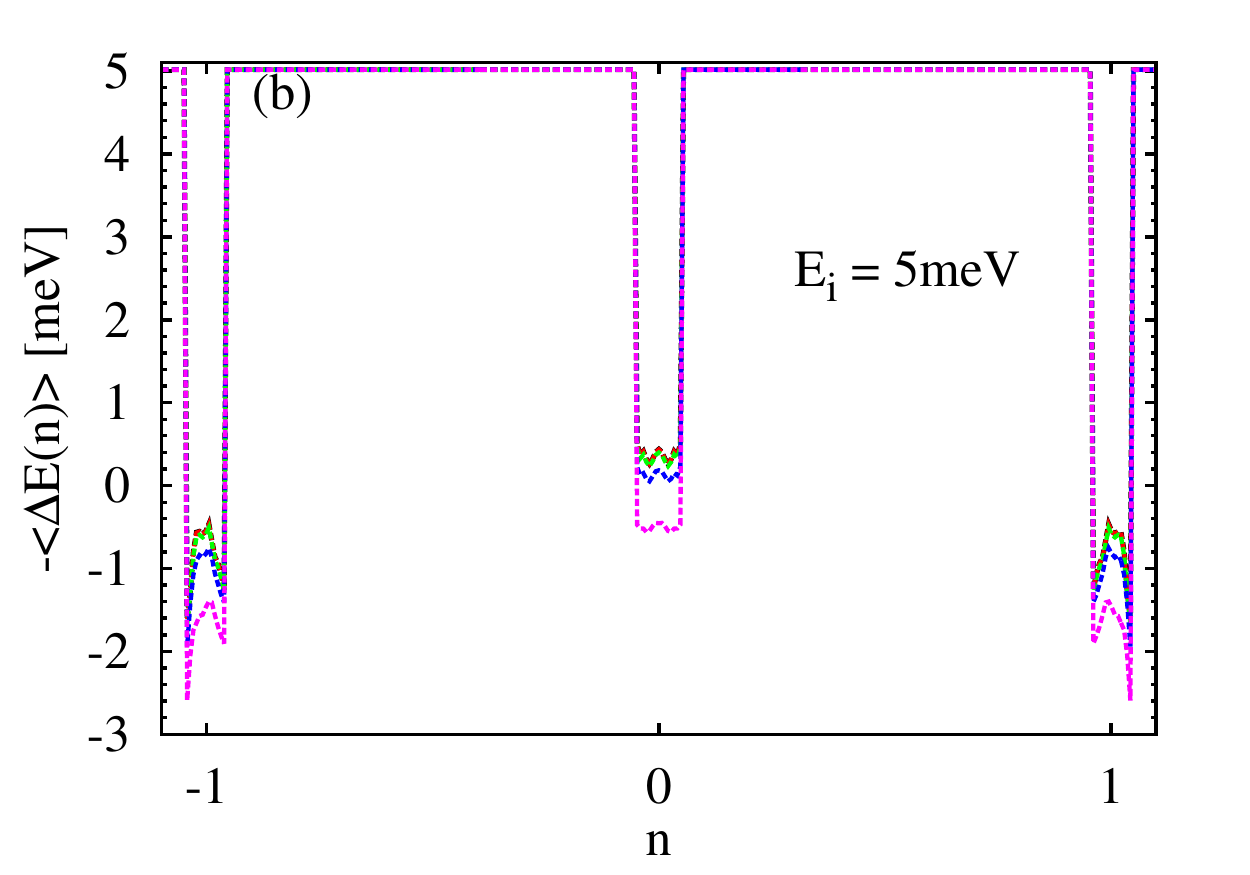}\tabularnewline
\end{tabular}
\caption{Same as Figure \ref{fig:fig9}, but the quantities are obtained by
performing MCTDH simulations.}
\label{fig:fig10}
\end{figure}

Figure \ref{fig:fig4} shows that the normalized energy loss obtained by Eq. (\ref{2.15}) 
as a function of diffraction quantum number. The normalized energy loss computed 
by classical simulation for inelastic scattering is higher (in negative magnitude) 
than the elastic scattering. Similar trend is followed by the quantum results though
energy loss is occurred around the diffraction quantum numbers, $n$. 

\begin{figure}[htp]
\centering
\begin{tabular}{r@{\extracolsep{0pt}.}lc}
\includegraphics[width=13cm,height=13cm,keepaspectratio]{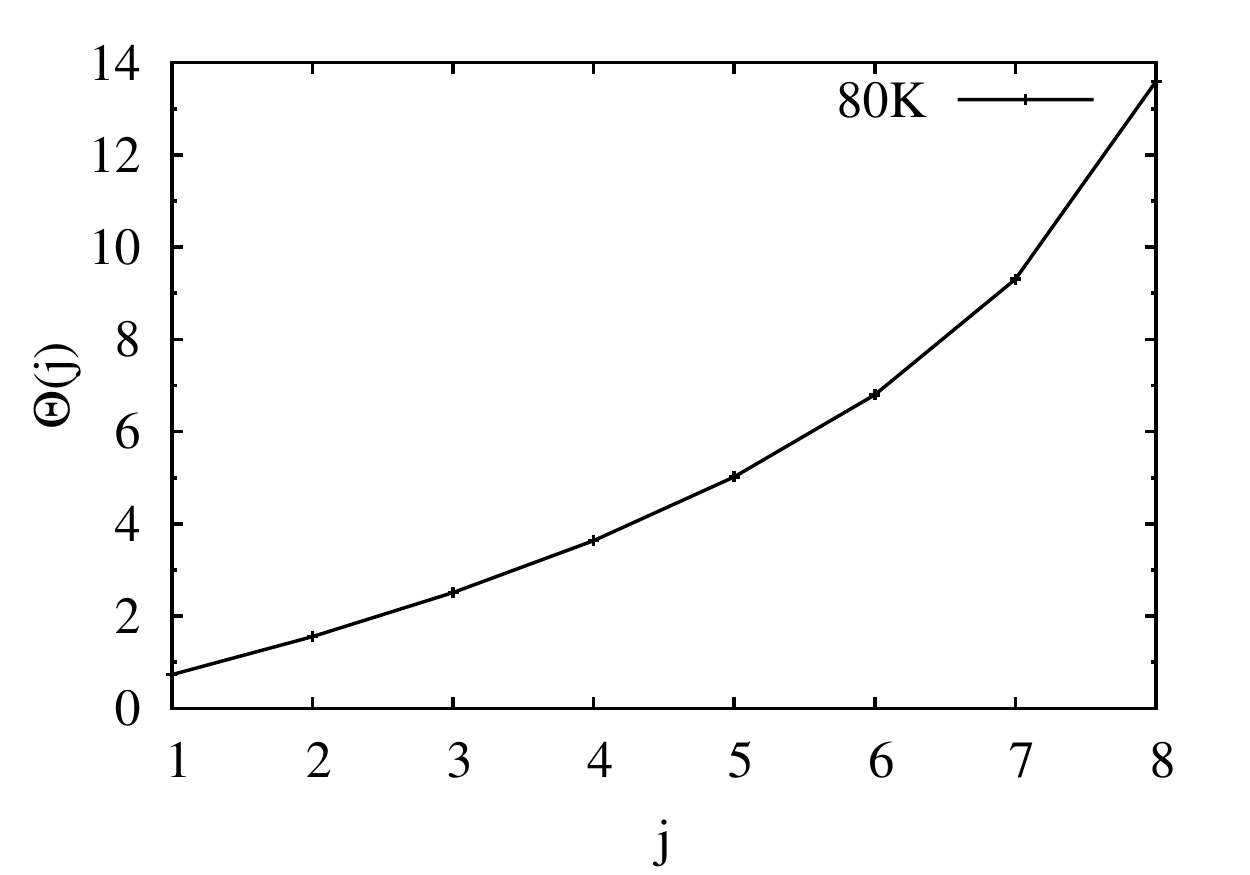}\tabularnewline 
\end{tabular}
\caption{The quantity, $\Theta (j) = \frac{\hbar \omega_{j}}{k_{B}T}$ as a function 
of index of bath.}
\label{fig:fig11}
\end{figure}

\renewcommand{\theequation}{4.\arabic{equation}} \setcounter{section}{3} %
\setcounter{equation}{0}

\section{Discussion}

We have performed classical and quantum dynamics on a model Hamiltonian
describing the scattering process on a thermal corrugated surface and 
compared the results at various temperatures for two incident energies of 
the particle. The model calculations displays that the classical trapping 
probability and the average energy of the escaped particle is higher than 
the quantum results at very low incident energies. Though the final angular
distributions computed classical as well as quantum dynamics are almost 
temperature independent, they are qualitatively different. Since our present 
study is limited to zero incidence angle, one can perform the dynamics 
at non-zero incidence angle for atom - surface scattering which is more 
realistic. 

\section*{Acknowledgment}

We thank Professor H.-D. Meyer for his help in employing the MCTDH Heidelberg 
package. 

\bibliographystyle{aipnum4-1}
\bibliography{article}

\begin{thebibliography}{40}%
\makeatletter
\providecommand \@ifxundefined [1]{%
 \@ifx{#1\undefined}
}%
\providecommand \@ifnum [1]{%
 \ifnum #1\expandafter \@firstoftwo
 \else \expandafter \@secondoftwo
 \fi
}%
\providecommand \@ifx [1]{%
 \ifx #1\expandafter \@firstoftwo
 \else \expandafter \@secondoftwo
 \fi
}%
\providecommand \natexlab [1]{#1}%
\providecommand \enquote  [1]{``#1''}%
\providecommand \bibnamefont  [1]{#1}%
\providecommand \bibfnamefont [1]{#1}%
\providecommand \citenamefont [1]{#1}%
\providecommand \href@noop [0]{\@secondoftwo}%
\providecommand \href [0]{\begingroup \@sanitize@url \@href}%
\providecommand \@href[1]{\@@startlink{#1}\@@href}%
\providecommand \@@href[1]{\endgroup#1\@@endlink}%
\providecommand \@sanitize@url [0]{\catcode `\\12\catcode `\$12\catcode
  `\&12\catcode `\#12\catcode `\^12\catcode `\_12\catcode `\%12\relax}%
\providecommand \@@startlink[1]{}%
\providecommand \@@endlink[0]{}%
\providecommand \url  [0]{\begingroup\@sanitize@url \@url }%
\providecommand \@url [1]{\endgroup\@href {#1}{\urlprefix }}%
\providecommand \urlprefix  [0]{URL }%
\providecommand \Eprint [0]{\href }%
\providecommand \doibase [0]{http://dx.doi.org/}%
\providecommand \selectlanguage [0]{\@gobble}%
\providecommand \bibinfo  [0]{\@secondoftwo}%
\providecommand \bibfield  [0]{\@secondoftwo}%
\providecommand \translation [1]{[#1]}%
\providecommand \BibitemOpen [0]{}%
\providecommand \bibitemStop [0]{}%
\providecommand \bibitemNoStop [0]{.\EOS\space}%
\providecommand \EOS [0]{\spacefactor3000\relax}%
\providecommand \BibitemShut  [1]{\csname bibitem#1\endcsname}%
\let\auto@bib@innerbib\@empty
\bibitem [{\citenamefont {Estermann}\ and\ \citenamefont
  {Stern}(1930)}]{zp-61-95-1930}%
  \BibitemOpen
  \bibfield  {author} {\bibinfo {author} {\bibfnamefont {I.}~\bibnamefont
  {Estermann}}\ and\ \bibinfo {author} {\bibfnamefont {A.}~\bibnamefont
  {Stern}},\ }\href@noop {} {\bibfield  {journal} {\bibinfo  {journal} {Z.
  Phys.}\ }\textbf {\bibinfo {volume} {61}},\ \bibinfo {pages} {95} (\bibinfo
  {year} {1930})}\BibitemShut {NoStop}%
\bibitem [{\citenamefont {Barker}\ and\ \citenamefont
  {Auerbach}(1984)}]{ssr-4-1-1984}%
  \BibitemOpen
  \bibfield  {author} {\bibinfo {author} {\bibfnamefont {J.~A.}\ \bibnamefont
  {Barker}}\ and\ \bibinfo {author} {\bibfnamefont {D.~J.}\ \bibnamefont
  {Auerbach}},\ }\href@noop {} {\bibfield  {journal} {\bibinfo  {journal}
  {Surf. Sci. Rep.}\ }\textbf {\bibinfo {volume} {4}},\ \bibinfo {pages} {1}
  (\bibinfo {year} {1984})}\BibitemShut {NoStop}%
\bibitem [{\citenamefont {Smith}, \citenamefont {Kara},\ and\ \citenamefont
  {Holloway}(1991)}]{jcp-94-806-1991}%
  \BibitemOpen
  \bibfield  {author} {\bibinfo {author} {\bibfnamefont {R.~J.}\ \bibnamefont
  {Smith}}, \bibinfo {author} {\bibfnamefont {A.}~\bibnamefont {Kara}}, \ and\
  \bibinfo {author} {\bibfnamefont {S.}~\bibnamefont {Holloway}},\ }\href@noop
  {} {\bibfield  {journal} {\bibinfo  {journal} {J. Chem. Phys.}\ }\textbf
  {\bibinfo {volume} {94}},\ \bibinfo {pages} {806} (\bibinfo {year}
  {1991})}\BibitemShut {NoStop}%
\bibitem [{\citenamefont {Schlichting}\ \emph {et~al.}(1992)\citenamefont
  {Schlichting}, \citenamefont {Menzel}, \citenamefont {Brunner},\ and\
  \citenamefont {Brenig}}]{jcp-97-4453-1992}%
  \BibitemOpen
  \bibfield  {author} {\bibinfo {author} {\bibfnamefont {H.}~\bibnamefont
  {Schlichting}}, \bibinfo {author} {\bibfnamefont {D.}~\bibnamefont {Menzel}},
  \bibinfo {author} {\bibfnamefont {T.}~\bibnamefont {Brunner}}, \ and\
  \bibinfo {author} {\bibfnamefont {W.}~\bibnamefont {Brenig}},\ }\href@noop {}
  {\bibfield  {journal} {\bibinfo  {journal} {J. Chem. Phys.}\ }\textbf
  {\bibinfo {volume} {97}},\ \bibinfo {pages} {4453} (\bibinfo {year}
  {1992})}\BibitemShut {NoStop}%
\bibitem [{\citenamefont {Brivio}\ and\ \citenamefont
  {Grimley}(1993)}]{ssr-17-1-1993}%
  \BibitemOpen
  \bibfield  {author} {\bibinfo {author} {\bibfnamefont {G.~P.}\ \bibnamefont
  {Brivio}}\ and\ \bibinfo {author} {\bibfnamefont {T.~B.}\ \bibnamefont
  {Grimley}},\ }\href@noop {} {\bibfield  {journal} {\bibinfo  {journal} {Surf.
  Sci. Rep.}\ }\textbf {\bibinfo {volume} {17}},\ \bibinfo {pages} {1}
  (\bibinfo {year} {1993})}\BibitemShut {NoStop}%
\bibitem [{\citenamefont {Mullins}, \citenamefont {Rettner},\ and\
  \citenamefont {Auerbach}(1989)}]{cpl-163-111-1989}%
  \BibitemOpen
  \bibfield  {author} {\bibinfo {author} {\bibfnamefont {C.~B.}\ \bibnamefont
  {Mullins}}, \bibinfo {author} {\bibfnamefont {C.~T.}\ \bibnamefont
  {Rettner}}, \ and\ \bibinfo {author} {\bibfnamefont {D.~J.}\ \bibnamefont
  {Auerbach}},\ }\href@noop {} {\bibfield  {journal} {\bibinfo  {journal}
  {Chem. Phys. Lett.}\ }\textbf {\bibinfo {volume} {163}},\ \bibinfo {pages}
  {111} (\bibinfo {year} {1989})}\BibitemShut {NoStop}%
\bibitem [{\citenamefont {Rettner}, \citenamefont {Bethune},\ and\
  \citenamefont {Auerbach}(1989)}]{jcp-91-1942-1989}%
  \BibitemOpen
  \bibfield  {author} {\bibinfo {author} {\bibfnamefont {C.~T.}\ \bibnamefont
  {Rettner}}, \bibinfo {author} {\bibfnamefont {D.~S.}\ \bibnamefont
  {Bethune}}, \ and\ \bibinfo {author} {\bibfnamefont {D.~J.}\ \bibnamefont
  {Auerbach}},\ }\href@noop {} {\bibfield  {journal} {\bibinfo  {journal} {J.
  Chem. Phys.}\ }\textbf {\bibinfo {volume} {91}},\ \bibinfo {pages} {1942}
  (\bibinfo {year} {1989})}\BibitemShut {NoStop}%
\bibitem [{\citenamefont {Rettner}, \citenamefont {Schweizer},\ and\
  \citenamefont {Mullins}(1989)}]{jcp-90-3800-1989}%
  \BibitemOpen
  \bibfield  {author} {\bibinfo {author} {\bibfnamefont {C.~T.}\ \bibnamefont
  {Rettner}}, \bibinfo {author} {\bibfnamefont {E.~K.}\ \bibnamefont
  {Schweizer}}, \ and\ \bibinfo {author} {\bibfnamefont {C.~B.}\ \bibnamefont
  {Mullins}},\ }\href@noop {} {\bibfield  {journal} {\bibinfo  {journal} {J.
  Chem. Phys.}\ }\textbf {\bibinfo {volume} {90}},\ \bibinfo {pages} {3800}
  (\bibinfo {year} {1989})}\BibitemShut {NoStop}%
\bibitem [{\citenamefont {Kondo}\ \emph {et~al.}(2005)\citenamefont {Kondo},
  \citenamefont {Kato}, \citenamefont {Yamada}, \citenamefont {Yamamoto},\ and\
  \citenamefont {Kawai}}]{jcp-122-244713-2005}%
  \BibitemOpen
  \bibfield  {author} {\bibinfo {author} {\bibfnamefont {T.}~\bibnamefont
  {Kondo}}, \bibinfo {author} {\bibfnamefont {H.~S.}\ \bibnamefont {Kato}},
  \bibinfo {author} {\bibfnamefont {T.}~\bibnamefont {Yamada}}, \bibinfo
  {author} {\bibfnamefont {S.}~\bibnamefont {Yamamoto}}, \ and\ \bibinfo
  {author} {\bibfnamefont {M.}~\bibnamefont {Kawai}},\ }\href@noop {}
  {\bibfield  {journal} {\bibinfo  {journal} {J. Chem. Phys.}\ }\textbf
  {\bibinfo {volume} {122}},\ \bibinfo {pages} {244713} (\bibinfo {year}
  {2005})}\BibitemShut {NoStop}%
\bibitem [{\citenamefont {Kondo}\ \emph {et~al.}(2006)\citenamefont {Kondo},
  \citenamefont {Kato}, \citenamefont {Yamada}, \citenamefont {Yamamoto},\ and\
  \citenamefont {Kawai}}]{epjd-38-129-2006}%
  \BibitemOpen
  \bibfield  {author} {\bibinfo {author} {\bibfnamefont {T.}~\bibnamefont
  {Kondo}}, \bibinfo {author} {\bibfnamefont {H.~S.}\ \bibnamefont {Kato}},
  \bibinfo {author} {\bibfnamefont {T.}~\bibnamefont {Yamada}}, \bibinfo
  {author} {\bibfnamefont {S.}~\bibnamefont {Yamamoto}}, \ and\ \bibinfo
  {author} {\bibfnamefont {M.}~\bibnamefont {Kawai}},\ }\href@noop {}
  {\bibfield  {journal} {\bibinfo  {journal} {Eur. Phys. J. D}\ }\textbf
  {\bibinfo {volume} {38}},\ \bibinfo {pages} {129} (\bibinfo {year}
  {2006})}\BibitemShut {NoStop}%
\bibitem [{\citenamefont {Hubbard}\ and\ \citenamefont
  {Miller}(1984)}]{jcp-80-5827-1984}%
  \BibitemOpen
  \bibfield  {author} {\bibinfo {author} {\bibfnamefont {L.~M.}\ \bibnamefont
  {Hubbard}}\ and\ \bibinfo {author} {\bibfnamefont {W.~H.}\ \bibnamefont
  {Miller}},\ }\href@noop {} {\bibfield  {journal} {\bibinfo  {journal} {J.
  Chem. Phys.}\ }\textbf {\bibinfo {volume} {80}},\ \bibinfo {pages} {5827}
  (\bibinfo {year} {1984})}\BibitemShut {NoStop}%
\bibitem [{\citenamefont {Tully}(1981)}]{sc-111-461-1981}%
  \BibitemOpen
  \bibfield  {author} {\bibinfo {author} {\bibfnamefont {J.~C.}\ \bibnamefont
  {Tully}},\ }\href@noop {} {\bibfield  {journal} {\bibinfo  {journal} {Surf.
  Sci.}\ }\textbf {\bibinfo {volume} {111}},\ \bibinfo {pages} {461} (\bibinfo
  {year} {1981})}\BibitemShut {NoStop}%
\bibitem [{\citenamefont {Tully}(1990{\natexlab{a}})}]{sc-226-180-1990}%
  \BibitemOpen
  \bibfield  {author} {\bibinfo {author} {\bibfnamefont {J.~C.}\ \bibnamefont
  {Tully}},\ }\href@noop {} {\bibfield  {journal} {\bibinfo  {journal} {Surf.
  Sci.}\ }\textbf {\bibinfo {volume} {226}},\ \bibinfo {pages} {180} (\bibinfo
  {year} {1990}{\natexlab{a}})}\BibitemShut {NoStop}%
\bibitem [{\citenamefont {Head-Gordon}\ \emph {et~al.}(1991)\citenamefont
  {Head-Gordon}, \citenamefont {Tully}, \citenamefont {Rettner}, \citenamefont
  {Mullins},\ and\ \citenamefont {Auerbach}}]{jcp-94-1516-1991}%
  \BibitemOpen
  \bibfield  {author} {\bibinfo {author} {\bibfnamefont {M.}~\bibnamefont
  {Head-Gordon}}, \bibinfo {author} {\bibfnamefont {J.~C.}\ \bibnamefont
  {Tully}}, \bibinfo {author} {\bibfnamefont {C.~T.}\ \bibnamefont {Rettner}},
  \bibinfo {author} {\bibfnamefont {C.~B.}\ \bibnamefont {Mullins}}, \ and\
  \bibinfo {author} {\bibfnamefont {D.~J.}\ \bibnamefont {Auerbach}},\
  }\href@noop {} {\bibfield  {journal} {\bibinfo  {journal} {J. Chem. Phys.}\
  }\textbf {\bibinfo {volume} {94}},\ \bibinfo {pages} {1516} (\bibinfo {year}
  {1991})}\BibitemShut {NoStop}%
\bibitem [{\citenamefont {Tully}(1990{\natexlab{b}})}]{jcp-92-680-1990}%
  \BibitemOpen
  \bibfield  {author} {\bibinfo {author} {\bibfnamefont {J.~C.}\ \bibnamefont
  {Tully}},\ }\href@noop {} {\bibfield  {journal} {\bibinfo  {journal} {J.
  Chem. Phys.}\ }\textbf {\bibinfo {volume} {92}},\ \bibinfo {pages} {680}
  (\bibinfo {year} {1990}{\natexlab{b}})}\BibitemShut {NoStop}%
\bibitem [{\citenamefont {Fan}\ and\ \citenamefont
  {Manson}(2009{\natexlab{a}})}]{prb-79-045424-2009}%
  \BibitemOpen
  \bibfield  {author} {\bibinfo {author} {\bibfnamefont {G.}~\bibnamefont
  {Fan}}\ and\ \bibinfo {author} {\bibfnamefont {J.~R.}\ \bibnamefont
  {Manson}},\ }\href@noop {} {\bibfield  {journal} {\bibinfo  {journal} {Phys.
  Rev. B}\ }\textbf {\bibinfo {volume} {79}},\ \bibinfo {pages} {045424}
  (\bibinfo {year} {2009}{\natexlab{a}})}\BibitemShut {NoStop}%
\bibitem [{\citenamefont {Fan}\ and\ \citenamefont
  {Manson}(2009{\natexlab{b}})}]{jcp-130-064703-2009}%
  \BibitemOpen
  \bibfield  {author} {\bibinfo {author} {\bibfnamefont {G.}~\bibnamefont
  {Fan}}\ and\ \bibinfo {author} {\bibfnamefont {J.~R.}\ \bibnamefont
  {Manson}},\ }\href@noop {} {\bibfield  {journal} {\bibinfo  {journal} {J.
  Chem. Phys.}\ }\textbf {\bibinfo {volume} {130}},\ \bibinfo {pages} {064703}
  (\bibinfo {year} {2009}{\natexlab{b}})}\BibitemShut {NoStop}%
\bibitem [{\citenamefont {Pollak}, \citenamefont {Sengupta},\ and\
  \citenamefont {Miret-Art\'{e}s}(2008)}]{jcp-129-054107-2008}%
  \BibitemOpen
  \bibfield  {author} {\bibinfo {author} {\bibfnamefont {E.}~\bibnamefont
  {Pollak}}, \bibinfo {author} {\bibfnamefont {S.}~\bibnamefont {Sengupta}}, \
  and\ \bibinfo {author} {\bibfnamefont {S.}~\bibnamefont {Miret-Art\'{e}s}},\
  }\href@noop {} {\bibfield  {journal} {\bibinfo  {journal} {J. Chem. Phys.}\
  }\textbf {\bibinfo {volume} {129}},\ \bibinfo {pages} {054107} (\bibinfo
  {year} {2008})}\BibitemShut {NoStop}%
\bibitem [{\citenamefont {Pollak}\ and\ \citenamefont
  {Miret-Art\'{e}s}(2009)}]{jcp-130-194710-2009}%
  \BibitemOpen
  \bibfield  {author} {\bibinfo {author} {\bibfnamefont {E.}~\bibnamefont
  {Pollak}}\ and\ \bibinfo {author} {\bibfnamefont {S.}~\bibnamefont
  {Miret-Art\'{e}s}},\ }\href@noop {} {\bibfield  {journal} {\bibinfo
  {journal} {J. Chem. Phys.}\ }\textbf {\bibinfo {volume} {130}},\ \bibinfo
  {pages} {194710} (\bibinfo {year} {2009})}\BibitemShut {NoStop}%
\bibitem [{\citenamefont {Pollak}\ and\ \citenamefont
  {Miret-Art\'{e}s}(2010{\natexlab{a}})}]{jcp-132-049901-2010}%
  \BibitemOpen
  \bibfield  {author} {\bibinfo {author} {\bibfnamefont {E.}~\bibnamefont
  {Pollak}}\ and\ \bibinfo {author} {\bibfnamefont {S.}~\bibnamefont
  {Miret-Art\'{e}s}},\ }\href@noop {} {\bibfield  {journal} {\bibinfo
  {journal} {J. Chem. Phys.}\ }\textbf {\bibinfo {volume} {132}},\ \bibinfo
  {pages} {049901(E)} (\bibinfo {year} {2010}{\natexlab{a}})}\BibitemShut
  {NoStop}%
\bibitem [{\citenamefont {Pollak}\ and\ \citenamefont
  {Tatchen}(2009)}]{prb-80-115404-2009}%
  \BibitemOpen
  \bibfield  {author} {\bibinfo {author} {\bibfnamefont {E.}~\bibnamefont
  {Pollak}}\ and\ \bibinfo {author} {\bibfnamefont {J.}~\bibnamefont
  {Tatchen}},\ }\href@noop {} {\bibfield  {journal} {\bibinfo  {journal} {Phys.
  Rev. B}\ }\textbf {\bibinfo {volume} {80}},\ \bibinfo {pages} {115404}
  (\bibinfo {year} {2009})}\BibitemShut {NoStop}%
\bibitem [{\citenamefont {Pollak}\ and\ \citenamefont
  {Tatchen}(2010)}]{prb-81-049903-2010}%
  \BibitemOpen
  \bibfield  {author} {\bibinfo {author} {\bibfnamefont {E.}~\bibnamefont
  {Pollak}}\ and\ \bibinfo {author} {\bibfnamefont {J.}~\bibnamefont
  {Tatchen}},\ }\href@noop {} {\bibfield  {journal} {\bibinfo  {journal} {Phys.
  Rev. B}\ }\textbf {\bibinfo {volume} {81}},\ \bibinfo {pages} {049903
  (erratum)} (\bibinfo {year} {2010})}\BibitemShut {NoStop}%
\bibitem [{\citenamefont {Pollak}, \citenamefont {Moix},\ and\ \citenamefont
  {Miret-Art\'{e}s}(2009)}]{prb-80-165420-2009}%
  \BibitemOpen
  \bibfield  {author} {\bibinfo {author} {\bibfnamefont {E.}~\bibnamefont
  {Pollak}}, \bibinfo {author} {\bibfnamefont {J.}~\bibnamefont {Moix}}, \ and\
  \bibinfo {author} {\bibfnamefont {S.}~\bibnamefont {Miret-Art\'{e}s}},\
  }\href@noop {} {\bibfield  {journal} {\bibinfo  {journal} {Phys. Rev. B}\
  }\textbf {\bibinfo {volume} {80}},\ \bibinfo {pages} {165420} (\bibinfo
  {year} {2009})}\BibitemShut {NoStop}%
\bibitem [{\citenamefont {Pollak}, \citenamefont {Moix},\ and\ \citenamefont
  {Miret-Art\'{e}s}(2010)}]{prb-81-039902-2010}%
  \BibitemOpen
  \bibfield  {author} {\bibinfo {author} {\bibfnamefont {E.}~\bibnamefont
  {Pollak}}, \bibinfo {author} {\bibfnamefont {J.}~\bibnamefont {Moix}}, \ and\
  \bibinfo {author} {\bibfnamefont {S.}~\bibnamefont {Miret-Art\'{e}s}},\
  }\href@noop {} {\bibfield  {journal} {\bibinfo  {journal} {Phys. Rev. B}\
  }\textbf {\bibinfo {volume} {81}},\ \bibinfo {pages} {039902(E)} (\bibinfo
  {year} {2010})}\BibitemShut {NoStop}%
\bibitem [{\citenamefont {Pollak}\ and\ \citenamefont
  {Miret-Art\'{e}s}(2010{\natexlab{b}})}]{cp-375-337-2010}%
  \BibitemOpen
  \bibfield  {author} {\bibinfo {author} {\bibfnamefont {E.}~\bibnamefont
  {Pollak}}\ and\ \bibinfo {author} {\bibfnamefont {S.}~\bibnamefont
  {Miret-Art\'{e}s}},\ }\href@noop {} {\bibfield  {journal} {\bibinfo
  {journal} {Chem. Phys.}\ }\textbf {\bibinfo {volume} {375}},\ \bibinfo
  {pages} {337} (\bibinfo {year} {2010}{\natexlab{b}})}\BibitemShut {NoStop}%
\bibitem [{\citenamefont {Daon}, \citenamefont {Pollak},\ and\ \citenamefont
  {Miret-Art\'{e}s}(2012)}]{jcp-137-201103-2012}%
  \BibitemOpen
  \bibfield  {author} {\bibinfo {author} {\bibfnamefont {S.}~\bibnamefont
  {Daon}}, \bibinfo {author} {\bibfnamefont {E.}~\bibnamefont {Pollak}}, \ and\
  \bibinfo {author} {\bibfnamefont {S.}~\bibnamefont {Miret-Art\'{e}s}},\
  }\href@noop {} {\bibfield  {journal} {\bibinfo  {journal} {J. Chem. Phys.}\
  }\textbf {\bibinfo {volume} {137}},\ \bibinfo {pages} {201103} (\bibinfo
  {year} {2012})}\BibitemShut {NoStop}%
\bibitem [{\citenamefont {Miret-Art\'{e}s}, \citenamefont {Daon},\ and\
  \citenamefont {Pollak}(2012)}]{jcp-136-204707-2012}%
  \BibitemOpen
  \bibfield  {author} {\bibinfo {author} {\bibfnamefont {S.}~\bibnamefont
  {Miret-Art\'{e}s}}, \bibinfo {author} {\bibfnamefont {S.}~\bibnamefont
  {Daon}}, \ and\ \bibinfo {author} {\bibfnamefont {E.}~\bibnamefont
  {Pollak}},\ }\href@noop {} {\bibfield  {journal} {\bibinfo  {journal} {J.
  Chem. Phys.}\ }\textbf {\bibinfo {volume} {136}},\ \bibinfo {pages} {204707}
  (\bibinfo {year} {2012})}\BibitemShut {NoStop}%
\bibitem [{\citenamefont {Daon}\ and\ \citenamefont
  {Pollak}(2015)}]{jcp-142-174102-2015}%
  \BibitemOpen
  \bibfield  {author} {\bibinfo {author} {\bibfnamefont {S.}~\bibnamefont
  {Daon}}\ and\ \bibinfo {author} {\bibfnamefont {E.}~\bibnamefont {Pollak}},\
  }\href@noop {} {\bibfield  {journal} {\bibinfo  {journal} {J. Chem. Phys.}\
  }\textbf {\bibinfo {volume} {142}},\ \bibinfo {pages} {174102} (\bibinfo
  {year} {2015})}\BibitemShut {NoStop}%
\bibitem [{\citenamefont {Zhou}, \citenamefont {Pollak},\ and\ \citenamefont
  {Miret-Art\'{e}s}(2014)}]{jcp-140-024709-2014}%
  \BibitemOpen
  \bibfield  {author} {\bibinfo {author} {\bibfnamefont {Y.}~\bibnamefont
  {Zhou}}, \bibinfo {author} {\bibfnamefont {E.}~\bibnamefont {Pollak}}, \ and\
  \bibinfo {author} {\bibfnamefont {S.}~\bibnamefont {Miret-Art\'{e}s}},\
  }\href@noop {} {\bibfield  {journal} {\bibinfo  {journal} {J. Chem. Phys.}\
  }\textbf {\bibinfo {volume} {140}},\ \bibinfo {pages} {024709} (\bibinfo
  {year} {2014})}\BibitemShut {NoStop}%
\bibitem [{\citenamefont {Sahoo}\ and\ \citenamefont
  {Pollak}(2015)}]{jcp-143-064706-2015}%
  \BibitemOpen
  \bibfield  {author} {\bibinfo {author} {\bibfnamefont {T.}~\bibnamefont
  {Sahoo}}\ and\ \bibinfo {author} {\bibfnamefont {E.}~\bibnamefont {Pollak}},\
  }\href@noop {} {\bibfield  {journal} {\bibinfo  {journal} {J. Chem. Phys.}\
  }\textbf {\bibinfo {volume} {143}},\ \bibinfo {pages} {064706} (\bibinfo
  {year} {2015})}\BibitemShut {NoStop}%
\bibitem [{\citenamefont {Pollak}\ and\ \citenamefont
  {Miret-Art\'{e}s}(2015)}]{jpcc-119-14532-2015}%
  \BibitemOpen
  \bibfield  {author} {\bibinfo {author} {\bibfnamefont {E.}~\bibnamefont
  {Pollak}}\ and\ \bibinfo {author} {\bibfnamefont {S.}~\bibnamefont
  {Miret-Art\'{e}s}},\ }\href@noop {} {\bibfield  {journal} {\bibinfo
  {journal} {J. Phys. Chem. C}\ }\textbf {\bibinfo {volume} {119}},\ \bibinfo
  {pages} {14532} (\bibinfo {year} {2015})}\BibitemShut {NoStop}%
\bibitem [{\citenamefont {Azuri}\ and\ \citenamefont
  {Pollak}(2015)}]{jcp-143-014705-2015}%
  \BibitemOpen
  \bibfield  {author} {\bibinfo {author} {\bibfnamefont {A.}~\bibnamefont
  {Azuri}}\ and\ \bibinfo {author} {\bibfnamefont {E.}~\bibnamefont {Pollak}},\
  }\href@noop {} {\bibfield  {journal} {\bibinfo  {journal} {J. Chem. Phys.}\
  }\textbf {\bibinfo {volume} {143}},\ \bibinfo {pages} {014705} (\bibinfo
  {year} {2015})}\BibitemShut {NoStop}%
\bibitem [{\citenamefont {Worth}\ \emph {et~al.}()\citenamefont {Worth},
  \citenamefont {Beck}, \citenamefont {J{\"a}ckle},\ and\ \citenamefont
  {Meyer}}]{mctdh:package}%
  \BibitemOpen
  \bibfield  {author} {\bibinfo {author} {\bibfnamefont {G.~A.}\ \bibnamefont
  {Worth}}, \bibinfo {author} {\bibfnamefont {M.~H.}\ \bibnamefont {Beck}},
  \bibinfo {author} {\bibfnamefont {A.}~\bibnamefont {J{\"a}ckle}}, \ and\
  \bibinfo {author} {\bibfnamefont {H.-D.}\ \bibnamefont {Meyer}},\ }\href@noop
  {} {}\bibinfo {howpublished} {The {MCTDH} {P}ackage, {V}ersion 8.2, (2000).
  H.-D. Meyer, {V}ersion 8.3 (2002), {V}ersion 8.4 (2007). Current version:
  8.4.12 (2016). {S}ee http://mctdh.uni-hd.de}\BibitemShut {NoStop}%
\bibitem [{\citenamefont {Meyer}, \citenamefont {Manthe},\ and\ \citenamefont
  {Cederbaum}(1990)}]{mey90:73}%
  \BibitemOpen
  \bibfield  {author} {\bibinfo {author} {\bibfnamefont {H.-D.}\ \bibnamefont
  {Meyer}}, \bibinfo {author} {\bibfnamefont {U.}~\bibnamefont {Manthe}}, \
  and\ \bibinfo {author} {\bibfnamefont {L.~S.}\ \bibnamefont {Cederbaum}},\
  }\href@noop {} {\bibfield  {journal} {\bibinfo  {journal} {Chem.\ Phys.\
  Lett.}\ }\textbf {\bibinfo {volume} {165}},\ \bibinfo {pages} {73} (\bibinfo
  {year} {1990})}\BibitemShut {NoStop}%
\bibitem [{\citenamefont {Manthe}, \citenamefont {Meyer},\ and\ \citenamefont
  {Cederbaum}(1992)}]{man92:3199}%
  \BibitemOpen
  \bibfield  {author} {\bibinfo {author} {\bibfnamefont {U.}~\bibnamefont
  {Manthe}}, \bibinfo {author} {\bibfnamefont {H.-D.}\ \bibnamefont {Meyer}}, \
  and\ \bibinfo {author} {\bibfnamefont {L.~S.}\ \bibnamefont {Cederbaum}},\
  }\href@noop {} {\bibfield  {journal} {\bibinfo  {journal} {J.~Chem.\ Phys.}\
  }\textbf {\bibinfo {volume} {97}},\ \bibinfo {pages} {3199} (\bibinfo {year}
  {1992})}\BibitemShut {NoStop}%
\bibitem [{\citenamefont {Beck}\ \emph {et~al.}(2000)\citenamefont {Beck},
  \citenamefont {J{\"a}ckle}, \citenamefont {Worth},\ and\ \citenamefont
  {Meyer}}]{bec00:1}%
  \BibitemOpen
  \bibfield  {author} {\bibinfo {author} {\bibfnamefont {M.~H.}\ \bibnamefont
  {Beck}}, \bibinfo {author} {\bibfnamefont {A.}~\bibnamefont {J{\"a}ckle}},
  \bibinfo {author} {\bibfnamefont {G.~A.}\ \bibnamefont {Worth}}, \ and\
  \bibinfo {author} {\bibfnamefont {H.-D.}\ \bibnamefont {Meyer}},\ }\href@noop
  {} {\bibfield  {journal} {\bibinfo  {journal} {Phys.\ Rep.}\ }\textbf
  {\bibinfo {volume} {324}},\ \bibinfo {pages} {1} (\bibinfo {year}
  {2000})}\BibitemShut {NoStop}%
\bibitem [{\citenamefont {Meyer}\ and\ \citenamefont
  {Worth}(2003)}]{mey03:251}%
  \BibitemOpen
  \bibfield  {author} {\bibinfo {author} {\bibfnamefont {H.-D.}\ \bibnamefont
  {Meyer}}\ and\ \bibinfo {author} {\bibfnamefont {G.~A.}\ \bibnamefont
  {Worth}},\ }\href@noop {} {\bibfield  {journal} {\bibinfo  {journal} {Theor.\
  Chem.\ Acc.}\ }\textbf {\bibinfo {volume} {109}},\ \bibinfo {pages} {251}
  (\bibinfo {year} {2003})}\BibitemShut {NoStop}%
\bibitem [{\citenamefont {Meyer}, \citenamefont {Gatti},\ and\ \citenamefont
  {Worth}(2009)}]{mey09:book}%
  \BibitemOpen
  \bibinfo {editor} {\bibfnamefont {H.-D.}\ \bibnamefont {Meyer}}, \bibinfo
  {editor} {\bibfnamefont {F.}~\bibnamefont {Gatti}}, \ and\ \bibinfo {editor}
  {\bibfnamefont {G.~A.}\ \bibnamefont {Worth}},\ eds.,\ \href@noop {} {\emph
  {\bibinfo {title} {{Multidimensional Quantum Dynamics: MCTDH Theory and
  Applications}}}}\ (\bibinfo  {publisher} {Wiley-VCH},\ \bibinfo {address}
  {Weinheim},\ \bibinfo {year} {2009})\BibitemShut {NoStop}%
\bibitem [{\citenamefont {Wang}, \citenamefont {Thoss},\ and\ \citenamefont
  {Miller}(2000)}]{jcp-112-47-2000}%
  \BibitemOpen
  \bibfield  {author} {\bibinfo {author} {\bibfnamefont {H.}~\bibnamefont
  {Wang}}, \bibinfo {author} {\bibfnamefont {M.}~\bibnamefont {Thoss}}, \ and\
  \bibinfo {author} {\bibfnamefont {W.~H.}\ \bibnamefont {Miller}},\
  }\href@noop {} {\bibfield  {journal} {\bibinfo  {journal} {J. Chem. Phys.}\
  }\textbf {\bibinfo {volume} {112}},\ \bibinfo {pages} {47} (\bibinfo {year}
  {2000})}\BibitemShut {NoStop}%
\bibitem [{\citenamefont {Rapaport}(1997)}]{leapfrog}%
  \BibitemOpen
  \bibfield  {author} {\bibinfo {author} {\bibfnamefont {D.~C.}\ \bibnamefont
  {Rapaport}},\ }\href@noop {} {\emph {\bibinfo {title} {The Art of Molecular
  Dynamics Simulation}}}\ (\bibinfo  {publisher} {Cambridge University Press},\
  \bibinfo {address} {Cambridge, UK},\ \bibinfo {year} {1997})\BibitemShut
  {NoStop}%
\end{thebibliography}%
\end{document}